\documentclass[fleqn,10pt]{wlscirep}
\usepackage[utf8]{inputenc}
\usepackage[T1]{fontenc}
\usepackage{amsmath}
\usepackage{soul}
\usepackage{caption}
\usepackage{float}
\usepackage{graphicx}
\usepackage{subfigure}
\usepackage{epstopdf} 
\usepackage[export]{adjustbox}
\usepackage[justification=justified]{caption}

\title{Deep Learning Reveals Underlying Physics of Light-matter Interactions in Nanophotonic Devices}
\author[1,a]{Yashar Kiarashinejad}
\author[1,a]{Sajjad Abdollahramezani}
\author[1]{Mohammadreza Zandehshahvar}
\author[1]{Omid Hemmatyar}

\author[1,b]{Ali Adibi}
\affil[1]{School of Electrical and Computer Engineering, Georgia Institute of Technology, 778 Atlantic Drive NW, Atlanta, GA 30332, USA}

\affil[a]{These authors contributed equally to this work}
\affil[b]{ali.adibi@ece.gatech.edu}

\begin{abstract}
In this paper, we present a deep learning-based (DL-based) algorithm, as a purely mathematical platform, for providing intuitive understanding of the properties of electromagnetic (EM) wave-matter interaction in nanostructures. This approach is based on using the dimensionality reduction (DR) technique to significantly reduce the dimensionality of a generic EM wave-matter interaction problem without imposing significant error. Such an approach implicitly provides useful information about the role of different features (or design parameters such as geometry) of the nanostructure in its response functionality. To demonstrate the practical capabilities of this DL-based technique, we apply it to a reconfigurable optical metadevice enabling dual-band and triple-band optical absorption in the telecommunication window. Combination of the proposed approach with existing commercialized full-wave simulation tools offers a powerful toolkit to extract basic mechanisms of wave-matter interaction in complex EM devices and facilitate the design and optimization of nanostructures for a large range of applications including imaging, spectroscopy, and signal processing. It is worth to mention that the demonstrated approach is general and can be used in a large range of problems as long as enough training data can be provided. 

\end{abstract}
\begin{document}

\flushbottom
\maketitle

\thispagestyle{empty}

\noindent \textbf{Keywords}: deep learning, physical understanding, dimensionality reduction, nanophotonics, metamaterials, plasmonics

\section{Introduction} \label{Intro}

To manipulate the inherent properties (e.g., amplitude, phase, polarization, and frequency) of electromagnetic (EM) waves in the subwavelength regime, nanophotonic structures (especially metamaterials and metasurfaces) have emerged as a promising candidate \cite{jahani2016all, yu2011light, arbabi2015dielectric, hsiao2017fundamentals}. The development of reliable fabrication techniques to realize such nanostructures has opened up new opportunities for forming reliable flat optical components to replace the existing bulky optical elements. Numerous interesting functionalities have been demonstrated so far including planar lenses \cite{ khorasaninejad2016metalenses}, calculus metasurfaces \cite{abdollahramezani2015analog, chizari2016analog, abdollahramezani2017dielectric,campbell2019review,sakurai2019ultranarrow,pestourie2018inverse,ma2018deep}, meta-holograms \cite{chen2013high}, and nonlinear meta-modulators \cite{taghinejad2018ultrafast, taghinejad2018hot}. However, systematic realization of mature optical functionalities using complex nanostructures requires significant knowledge about the influence of nanostructure features on the interaction of EM waves,  which currently can only be found using cumbersome numerical calculations. Despite extensive efforts in forming new approaches for design and optimization of nanostructures \cite{campbell2019review}(e.g., using brute-force techniques \cite{Seidel1994}, evolutionary techniques like genetic algorithms \cite{Gondarenko2008,Hakansson2005} or particle swarm optimization \cite{ong2017freestanding}, semi-analytical modeling \cite{Piggott2017,Lu2013,Su2018,Frellsen2016,Piggott2014,Englund2005,molesky2018inverse,Mansouree:18,hemmatyar2017phase}, pattern recognition method\cite{melati2018mapping}, and even neural network-based techniques \cite{Ma2018,baxter2019plasmonic,,Liu2018a,Peurifoy2018,Liu2018,Tahersima2018,Zhang2018a,Ma2018,Qu2018,Inampudi2018,yao2018intelligent,so2019simultaneous,asano2018optimization,Zhang2018a}), systematic approaches for understanding the physics of wave-matter interaction in nanostructures and/or the effect of structural properties on their output response are still missing. On the other hand, available design and optimization approaches (e.g., the brute-force techniques) either suffer from significant computation complexity or over-simplify the problem in both response and design domains (e.g., due to sever down-sampling). Such computation complexity and oversimplification hinders the use of the existing design tools to provide a detailed understanding of the dynamics of light-matter interaction inside nanostructures unless a large set of simulations is performed. 

Here, we present a new design and optimization approach based on deep learning (DL) that provides detailed information about the role of design parameters in the output response of any nanostructure as well as intuitive understanding of the physics of wave-matter interaction in these nanostructures without imposing stringent computation complexity. Our approach is based on reducing the dimensionality of the problem in both design and response spaces while preserving the vital information. By training proper neural networks (NNs) for implementation of these dimensionality reduction (DR) processes, we find a complex analytic formula that relates the design parameters to the output response of the nanostructure. Despite their inherent complexity, such analytic relationships provide valuable intuitive information about the role of each design parameter in the overall response of the nanostructure at minimal computation costs. This is in contrast to existing design and optimization techniques, which require costly iterative computations for each design problem without providing intuitive understanding about the role of design parameters \cite{peurifoy2018nanophotonic,liu2018generative,qu2018migrating}.  This approach also allows for trading off the computation accuracy and complexity. Thus, it can be used for obtaining quick high-level information about the physics of light-matter interaction (e.g., the role of a given design parameter on the overall device performance) or running longer simulations to achieve more detailed information about a specific feature of the structure. By providing the required information for intuitive understanding of the light-matter interaction or design of a class of patterned nanostructures for any desired application with orders of magnitude less computation complexity, this approach can have a transformative impact on several applications that rely on nanostructures including imaging, spectroscopy, signal analysis, sensing, and LiDAR among others. 

The design and optimization approach and its important properties are discussed in Section \ref{Design}. The results of the application of this technique to practical nanostructures is presented in Section \ref{Analysis}. Understanding the underlying physics of investigated nanostructures as well as more detailed properties of this technique is discussed in Sections \ref{Understanding} and \ref{Discus}, and final conclusions are made in Section \ref{Conclusion}. 

\section{Deep learning-based Approach for Design and Optimization of Nanostructures} \label{Design}

\begin{figure}
    \centering
    \includegraphics[width=1\textwidth, trim={0cm 2cm 0cm 2cm},clip]{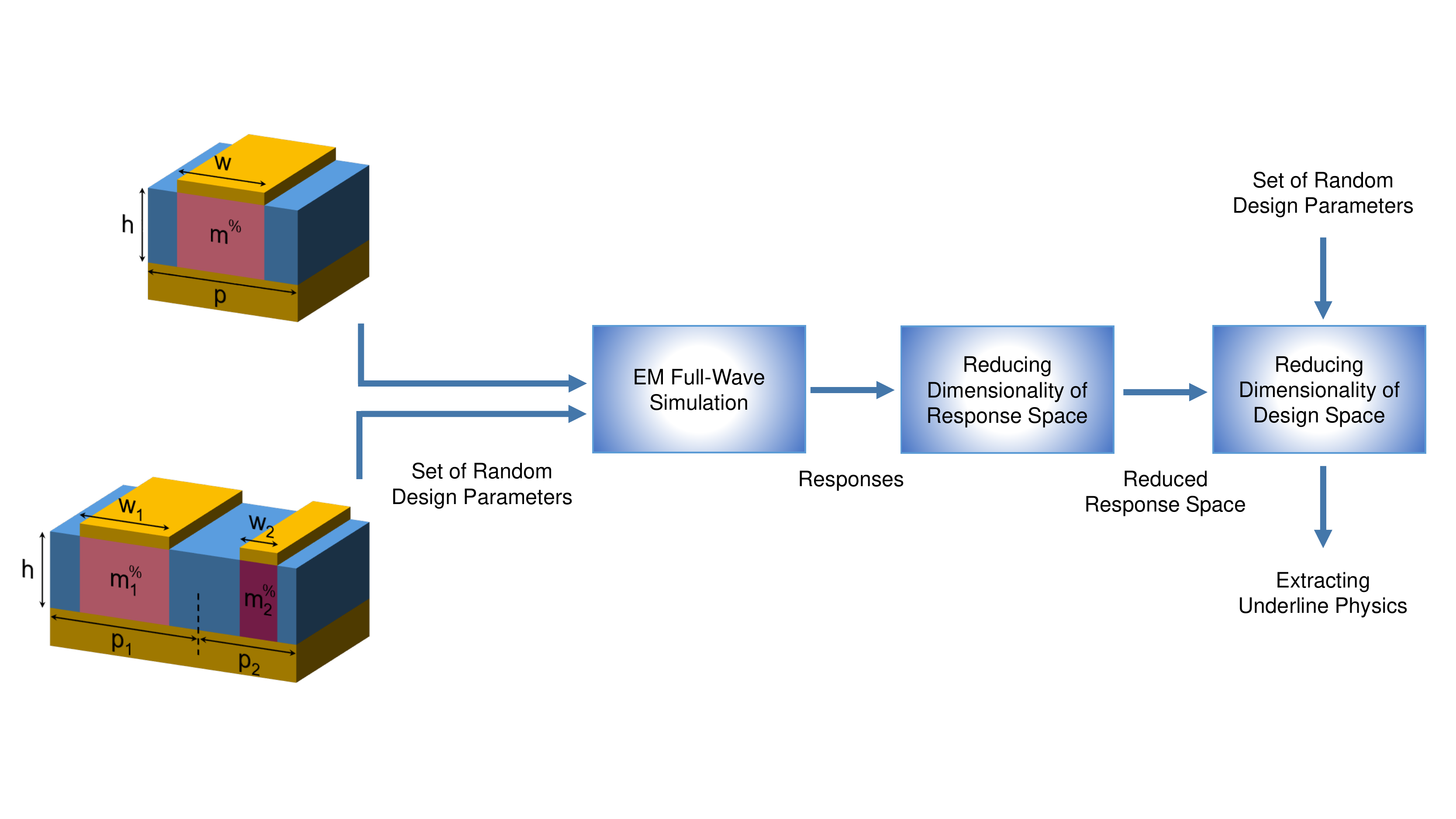}
    \caption{Procedure of revealing the underlying physics of a generic nanostructure using the DR algorithm. First, a set of random values is fed into the full-wave EM solver as design parameters. The output response (e.g., reflection spectra) of the corresponding simulated structures are lunched to a DR algorithm forming the reduced response space. In the next step, the reduced response-space data points and the associated random design parameters are used for training of a pseudo-encoder for DR of the design space. Finally, the trained pseudo-encoder provides physical intuition of the investigated EM structure and the role of each design parameter on the overall output response.}
    \label{fig:fig1_blkdiagram}
\end{figure}

Figure \ref{fig:fig1_blkdiagram} shows the schematic representation of our platform for analysis, design, optimization, and understanding the physics of nanostructures. Our main focus in this paper is to use such a simulation platform to extract the underlying physics of light-matter interaction through running a complete design and optimization process. Our approach (depicted in Fig. \ref{fig:fig1_blkdiagram}) uses the high level of correlation of light-matter interaction in the spatial and spectral domains to considerably reduce the dimensionality of the response space of the problem. Furthermore, the correlation that often exists among the effect of structural design parameters on the response space is used to reduce the dimensionality of the design space.

In the first step, a full-wave EM simulation software (based on the finite element method (FEM) implemented in the COMSOL environment, unless otherwise stated) is used to provide sufficient number of randomly generated instances (or so-called the input dataset) to train our DR approach. Each instance is calculated using a given set of randomly selected design parameters (i.e., a point in the design space), and thus, it relates the design space to the response space. After feeding a DR network using a subset of the available training data, we reduce the dimensionality of the response space. By setting the level of acceptable error, we find the minimum accessible dimensionality of the reduced response space. These reduced features are related to the intact dimensions in the original response space through the analytic formulas provided by the dimensionality expansion methods.

\begin{figure}[t]
    \centering
    \includegraphics[width=0.6\textwidth]{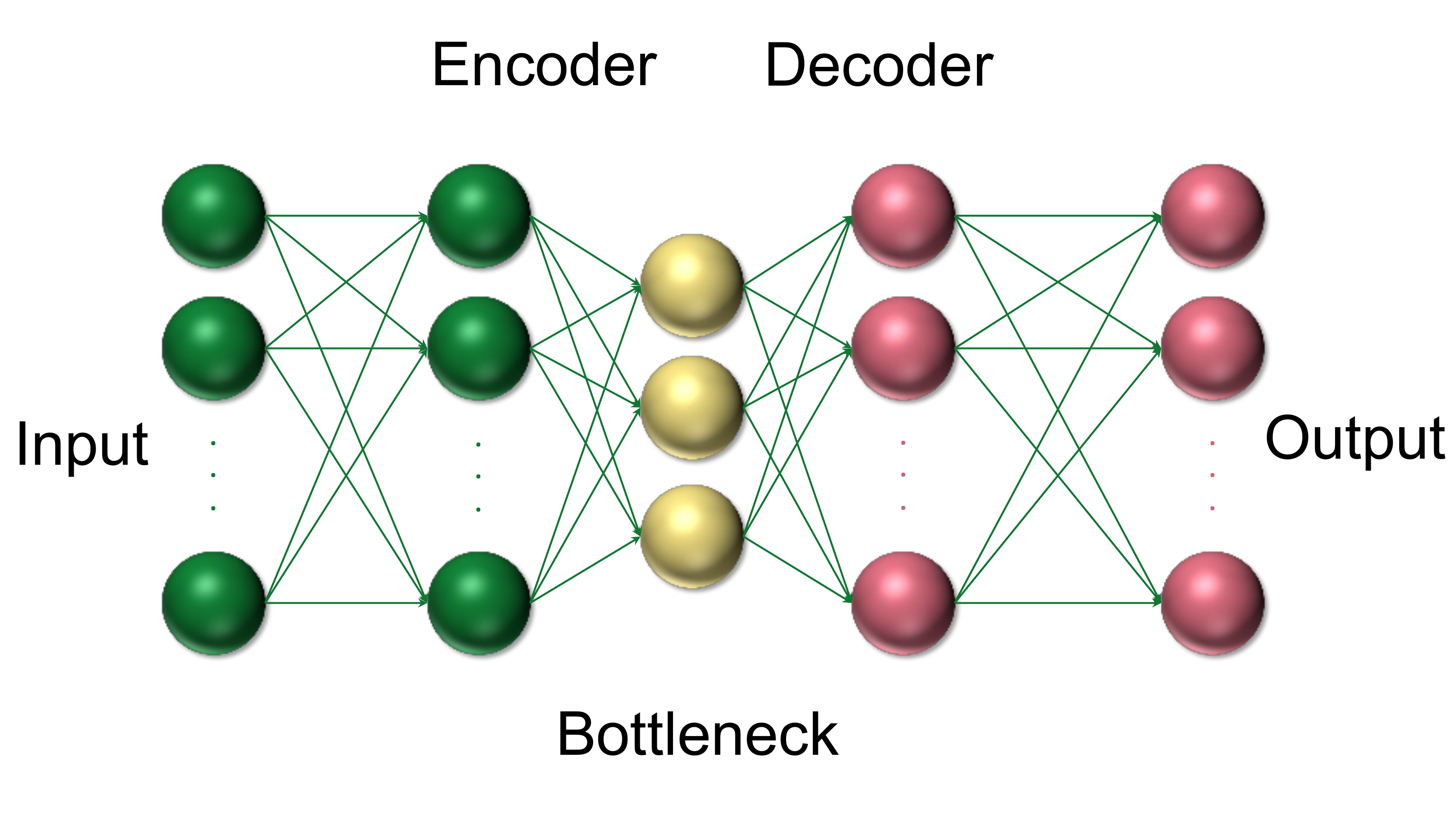}
    \caption{Schematic representation of an autoencoder architecture used in the DR technique. The leftmost half part (i.e., the encoder) reduces the dimensionality (the bottleneck layer represents the reduced space) while the right part (i.e., the decoder) recovers the data from the reduces space back to the original space. $x_{i}$ and $\hat{x}_{i}$ represent inputs and outputs, respectively.}
    \label{fig:fig2_AE}
\end{figure}

We applied principal component analysis (PCA) \cite{pearson1901liii}, kernel PCA (KPCA) \cite{scholkopf1997kernel}, and autoencoder \cite{Hinton2006} to reduce the dimensionality of our response space \cite{friedman2001elements} (more details are provided in the Supplementary Information). PCA is a linear DR algorithm that projects the data points on the eigenvectors of the covariance matrix of the responses. During the projection, first \textit{d} eigenvectors with the highest eigenvalues are selected, where \textit{d} represents the dimensionality of the reduced response space \cite{pearson1901liii}. KPCA is the nonlinear version of PCA in which a kernel function maps datapoints using a nonlinear function and then projects the datapoints on the basis vectors of the covariance matrix of the kernelized space \cite{scholkopf1997kernel}. As another DR method, the autoencoder (shown in Fig. \ref{fig:fig2_AE}) encodes the high-dimensional input on the leftmost part to low-dimensional data in the middle layer using a multilayer NN. The same NN can be used to decode and recover the data (back to the original response space) with some error \cite{kiarashinejad2019deep}. In other words, autoencoder is a feedforward NN in which it has same number of inputs and outputs. The number of neurons in the middle layer represents the dimension of the low-dimensional data represent the desired reduced dimensionality. This layer is also known as the bottleneck of the autoencoder. As shown in Section \ref{Analysis}, our simulations show that the performance of the autoencoder surpasses those of the PCA and KPCA. 

\begin{figure}[t]
    \centering
    \includegraphics[width=0.7\textwidth]{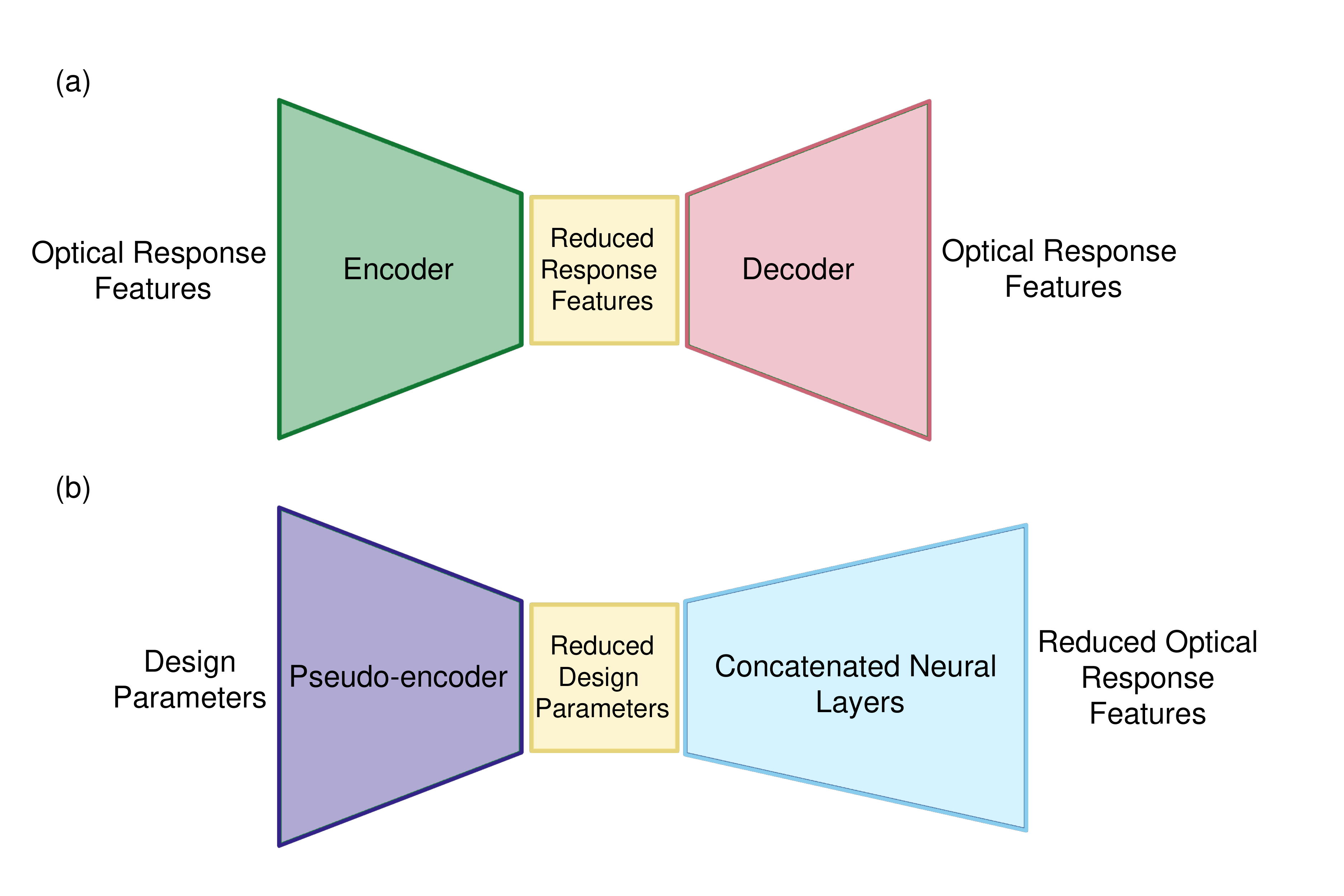}
    \caption{Reducing dimensionality of response and design spaces using autoencoder and pseudo-encoder platforms, respectively. (a) Reducing the optical response feature and represent optical response in reduced response space. (b) Architecture of a pseudo-encoder, that maps the design parameters to the reduced design parameters.}
    \label{fig:fig3_pseudoenc}
\end{figure}

After training an autoencoder for the DR of the response space (see Fig. \ref{fig:fig3_pseudoenc}(a)), we reduce the dimensionality of the design space using a pseudo-encoder architecture (see Fig. \ref{fig:fig3_pseudoenc}(b)) \cite{kiarashinejad2019deep} to relate the design and response spaces with minimal computation complexity. Since the input and output of the DR mechanism in Fig. \ref{fig:fig3_pseudoenc}(b) are different parameters (in contrast to the case for an autoencoder), we call this architecture a pseudo-encoder. In this manner, we directly include the information about the response of the nanostructure into the reduced design space, which is a major advantage of the pseudo-encoder architecture. Once the DR in both spaces are performed, we form a complete NN-based architecture that directly relates the design parameters to the nanostructure response by integrating the two trained NNs for the DR algorithms shown in Figs. \ref{fig:fig3_pseudoenc}(a) and \ref{fig:fig3_pseudoenc}(b). Once the underlying NNs are trained, we will obtain complex analytic formulas to study in details the roles of the design parameters in the output features. In addition, the weights of the NNs at different layers in for both the autoencoder and the pseudo-encoder can provide valuable information about the role of design parameters on the output response.

\section{Analysis of Nanostructures Using the DR-based Technique} \label{Analysis}

\begin{figure}[t]
    \centering
    \includegraphics[width=0.9\textwidth]{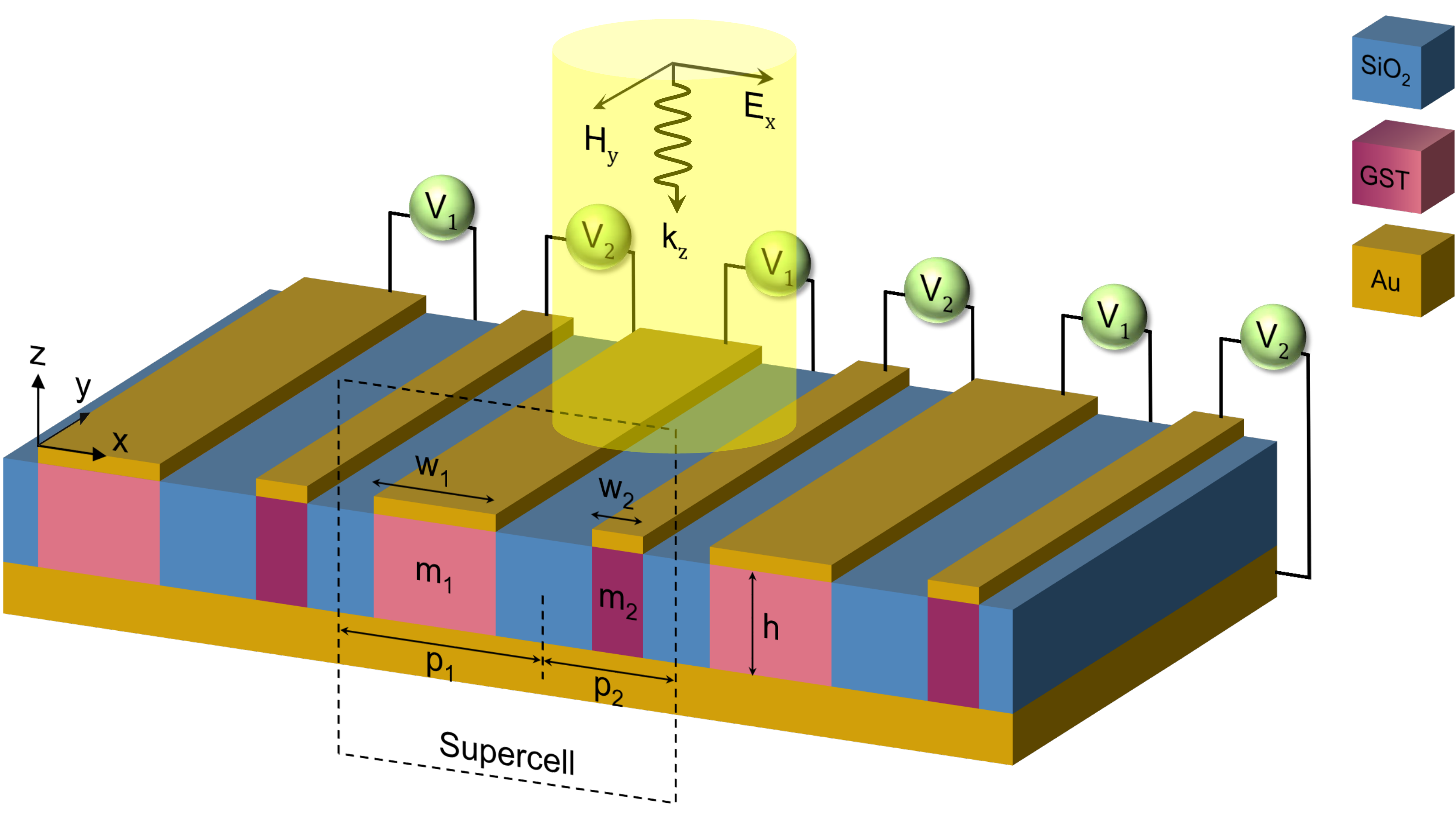}
    \caption{Three-dimensional (3D) illustration of the hybrid plasmonic/phase-change material metasurface studied in this paper. The design parameters are the thickness of GST nanostripes ($h$), the crystallization fraction of the GST nanostripes ($m_1^{\%}$ and $m_2^{\%}$), the unit cell periodicities ($p_1$ and $p_2$), and Au nanoribbon widths ($w_1$ and $w_2$) while the thickness of the Au nanoribbon is fixed at 30 nm. Each supercell can consist of one or two unit cells each comprised of a GST nanostripe encapsulated between the Au nanoribbon and the Au back-reflector separated from each other by symmetric SiO$_2$ spacers. $m_1^{\%}$ and $m_2^{\%}$ are dependent on the gate voltages $V_1$ and $V_2$, respectively. The whole structure is illuminated with a TM-polarized light at the near-infrared frequency range. }
    \label{fig:fig4_supercell}
\end{figure}

To show the applicability of the proposed approach, we consider here two simple design problems for the implementation of a reconfigurable multifunctional metadevice enabling dual-band and triple-band absorption in the telecommunication window. Figure \ref{fig:fig4_supercell} shows the schematic of the supercell structure of the metadevice, which can be electrically tuned to obtain the desired reflection spectrum when illuminated with a TM-polarized light (i.e., magnetic field normal to the direction of grating). Considering the maximum sampling value for the periodicity, the supercell in this design can consist of up to two unit cells to effectively suppress the higher diffraction orders in the telecommunication window. Each unit cell is comprised of a gold (Au) nanoribbon incorporating germanium antimony telluride (GST), a well-developed phase-change alloy. Upon non-volatile conversion of GST from the amorphous to the full crystalline state, a drastic change happens in its refractive index, which consequently induces a remarkable change in the reflection response. Meanwhile, the intermediate phase transition of GST can be realized by exciting it with an external stimulus (e.g., an electrical current). The nanostruture in Fig. \ref{fig:fig4_supercell} has 7 design parameters, i.e., the widths of the two Au nanoribbons in the supercell ($w_1$ and $w_2$), unit cell periodicities ($p_1$ and $p_2$), the crystalline states of the two GST nanostripes ($m_1^{\%}$ and $m_2^{\%}$ in which the superscript number represents the crystallization fraction), and the thickness of the GST nanostripes ($h$). It is notable that a specific crystallization fraction (which is associated to the refractive index of GST) can be realized by applying a predefined gate voltage. Here, we assume the thicknesses of the silicon dioxide (SiO$_{2}$) layer and Au nanoribbons are defined by the fabrication limitations, and thus, we do not consider it as a separate design parameter. 

The training of the DR algorithm for the response space is performed with 1700 instances obtained using the FEM simulations for structures with randomly selected design parameters. To obtain these instances, the reflection spectrum (i.e., reflection as a function of frequency) of the structure in Fig. \ref{fig:fig4_supercell} is calculated and sampled over the 150-300 THz range (with 3.75 THz spacing between adjacent samples) to obtain the response-space results. Thus, the number of parameters in the design space is 7, and the number of samples in the response space is 40. Details of these simulations are provided in Methods. In addition, we simulated 300 extra structures (with randomly selected design parameters) to obtain the validation dataset.

\begin{figure}
\hspace{1cm}
    \centering
    \hspace{-1cm}
 \subfigure[]{\includegraphics[page=1,width=0.4\textwidth, trim={5cm 13cm 4cm 4cm},clip]{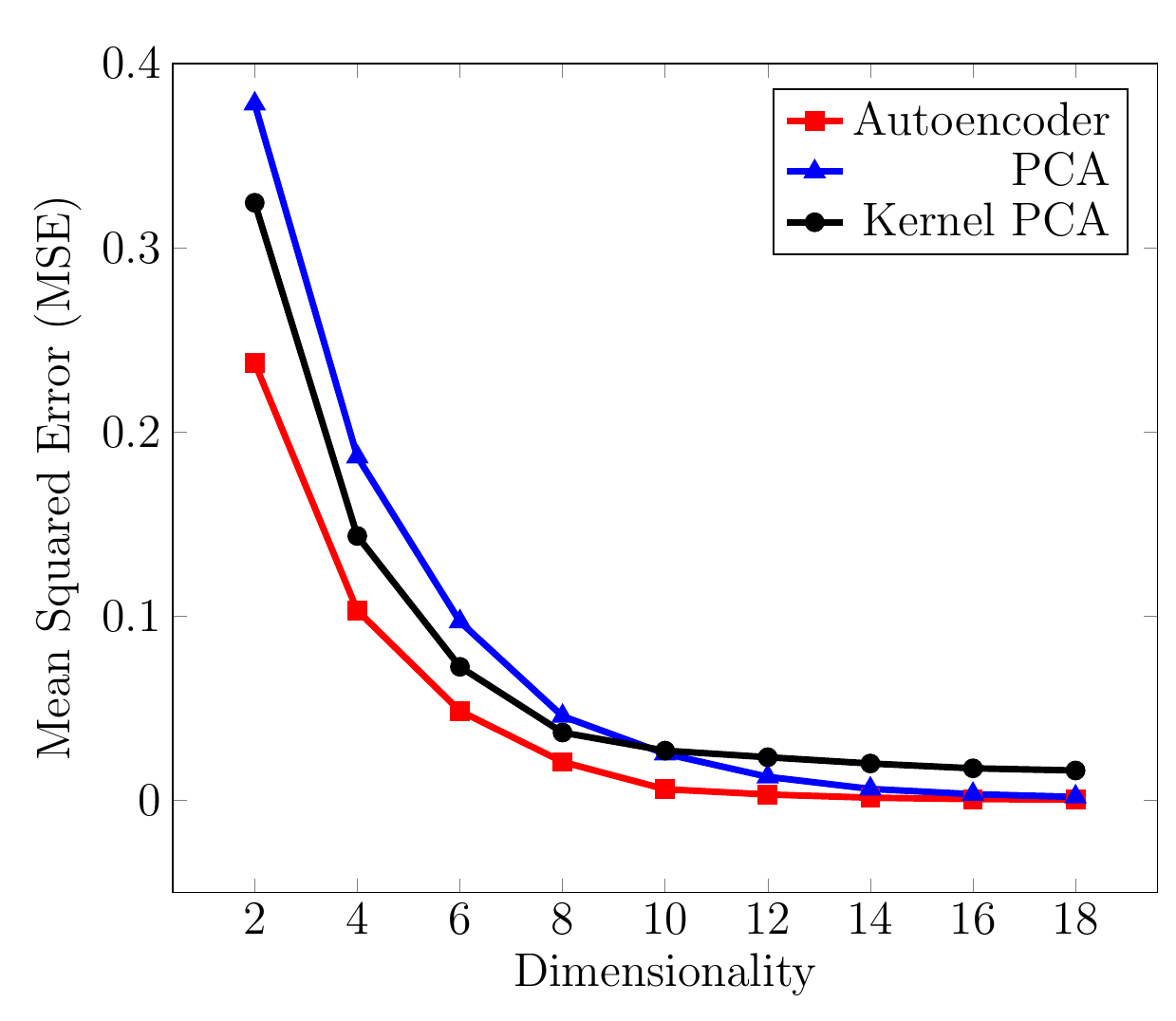}
    \hspace{1cm}
    \label{fig:Fig.6abb}}
    \hspace{-1cm}
    \subfigure[]{\includegraphics[page=2,width=0.4\textwidth, trim={5cm 12.7cm 4cm 4cm},clip]{Fig5_MSEDR.pdf}
    \hspace{1cm}
    \label{fig:fig5_msebb}}
    \caption{(a) MSE for different DR algorithms on the response space. The autoencoder has 7 layers with dimensions 40, 30, 20, \textit{d}, 20, 30, and 40, where \textit{d} represents the number of reduced dimensions. The KPCA is trained with a polynomial kernel of degree 7. (b) MSE of the pseudo-encoder for different dimensions of the reduced design space. The number of nodes (or the dimension) of the different layers of the  pseudo-encoder are 7, \textit{d}, 10, 20, 20, 30, 30, 40, and 7 at each layer where \textit{d} represents the dimensionality of the reduced design space.}
    \label{fig:fig5_mse}
    \end{figure}

\begin{figure}
    \centering
    \subfigure[]{\includegraphics[page=1,width=0.4\textwidth, trim={4cm 11cm 0cm 4cm},clip]{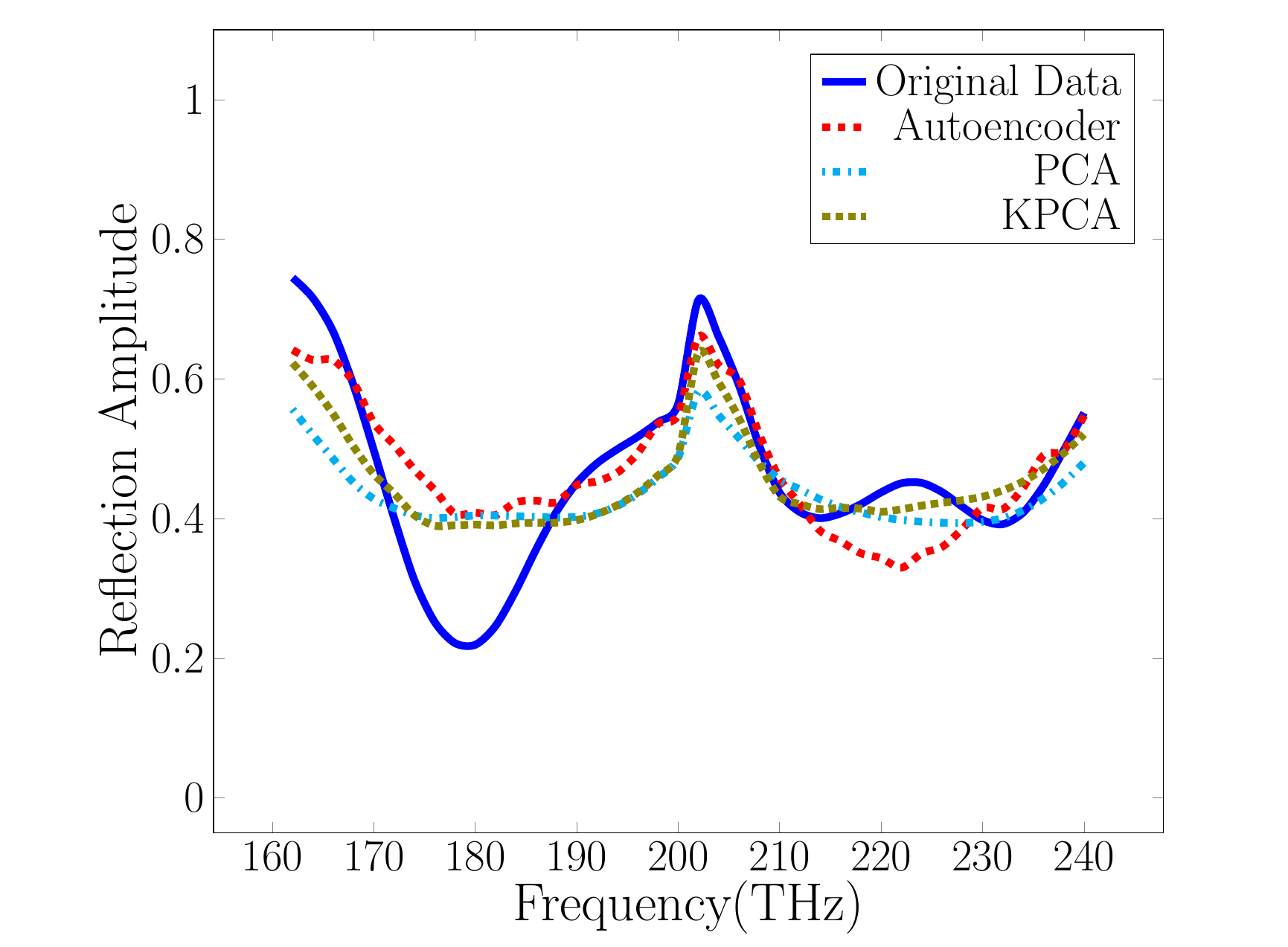}
   \label{fig:Fig.6a}}
    \subfigure[]{\includegraphics[page=1,width=0.4\textwidth, trim={3cm 10.5cm 1cm 3.2cm},clip]{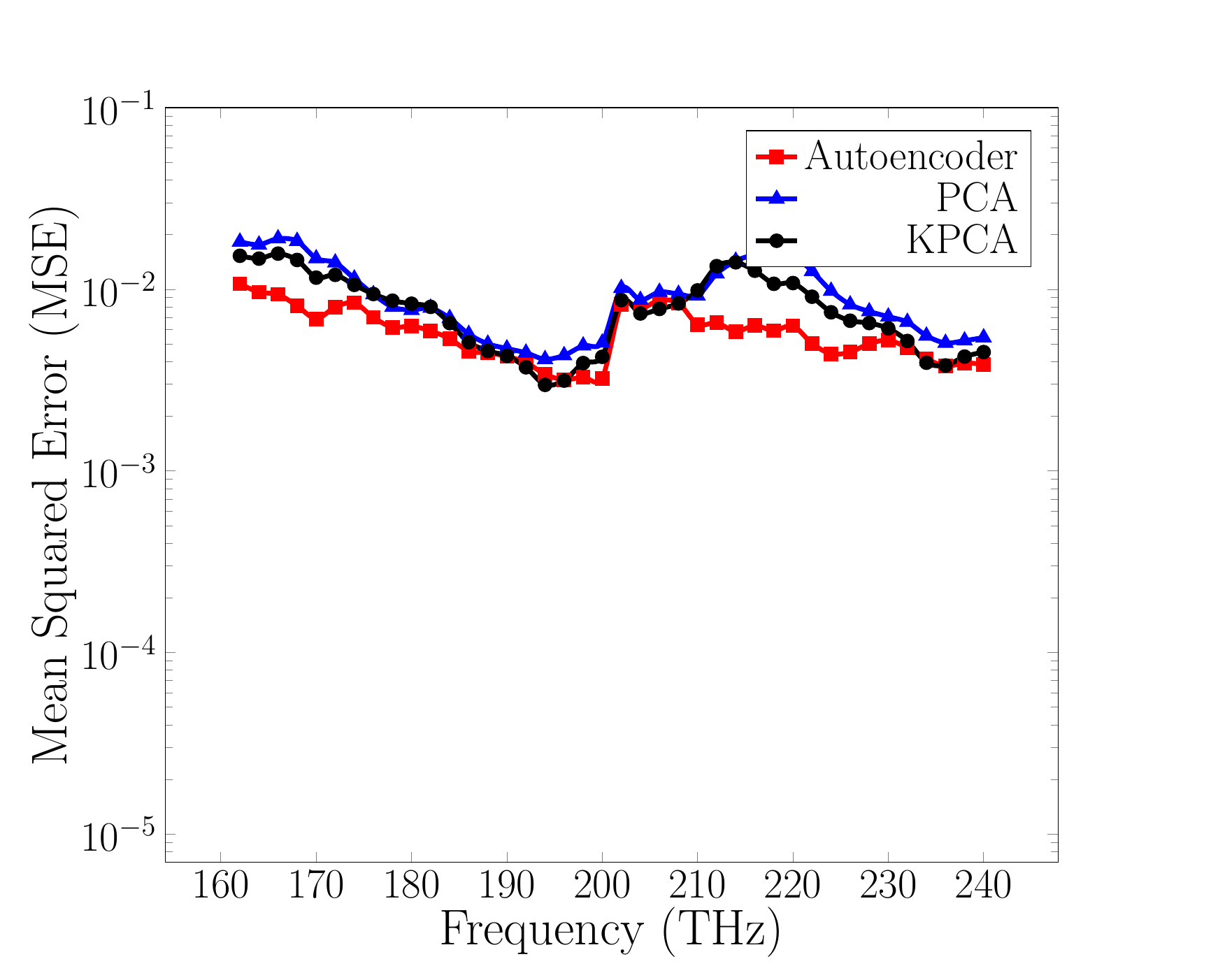}
    \label{fig:Fig.6b}}
    \subfigure[]{\includegraphics[page=2,width=0.4\textwidth, trim={4cm 11cm 0cm 4cm},clip]{Fig_6_Final.pdf}
    \label{fig:Fig.6c}}
    \subfigure[]{\includegraphics[page=2,width=0.4\textwidth, trim={3cm 10.5cm 1cm 3.2cm},clip]{Error_freq.pdf}
    \label{fig:Fig.6d}}
    \subfigure[]{\includegraphics[page=3,width=0.4\textwidth, trim={4cm 11cm 0cm 4cm},clip]{Fig_6_Final.pdf}
    \label{fig:Fig.6e}}
    \subfigure[]{\includegraphics[page=3,width=0.4\textwidth, trim={3cm 10.5cm 1cm 3.2cm},clip]{Error_freq.pdf}
    \label{fig:Fig.6f}}
    \caption{Reconstructed reflection amplitude (left) and  reconstruction MSE versus frequency (right) for dimensionality reduction using PCA, KPCA, and autoencoder. Results for reducing the dimensionality from 40 to (a),(b) 2, (c),(d) 7, and (e),(f) 16. The hyper-parameters for KPCA and autoencoder are the same as parameters used in fig. \ref{fig:fig5_mse}.}
    \label{fig:fig6_Rec}
\end{figure}

We applied the three DR algorithms (discussed on Section \ref{Design}) with different number of reduced dimensions to the training dataset and tested them with the (unseen) validation dataset. The mean squared error (MSE) for different number of dimensions (\textit{d}) in the reduced response space for the three DR algorithms is represented in Fig. \ref{fig:fig5_mse}(a). In these simulations, the polynomial kernel with degree of 7 is selected for the KPCA method. The autoencoder (see Fig. \ref{fig:fig2_AE}) consists of 7 layers in total and the number of nodes in the hidden (or intermediate) layers are 40, 30, 20, \textit{d}, 20, 30, and 40, respectively. Here \textit{d} represents the dimension of the reduced response space, which is the number of nodes in the bottleneck of the autoencoder in Fig. \ref{fig:fig2_AE}. The activation function for all nodes is fixed to tangent sigmoid function. As it is shown in Fig. \ref{fig:fig5_mse}(a), the autoencoder outperforms PCA and KPCA for all values of \textit{d}. This reveals the effectiveness of autoencoder in keeping nonlinear properties of the response space. KPCA works slightly better than PCA for lower dimensions (\textit{d}); however, it has higher MSE as the dimensionality increases because of overfitting. Figure \ref{fig:fig6_Rec} represents the reconstructed spectra using different DR methods for three values of \textit{d} (\textit{d} = 2, 7, and 16 for Figs. \ref{fig:fig6_Rec}(a), \ref{fig:fig6_Rec}(b), and \ref{fig:fig6_Rec}(c), respectively, with respective errors for different cases shown in Figs. \ref{fig:fig6_Rec}(d), \ref{fig:fig6_Rec}(e), and \ref{fig:fig6_Rec}(f)). As seen from Fig. \ref{fig:fig6_Rec}(b), the autoencoder is able to reconstruct the response spectrum after reducing the dimensionality of the response space from 40 to 7. Figure \ref{fig:fig5_mse}(a) also confirms that the autoencoder with \textit{d} = 7 is a good choice for the DR of the response space with MSE < 0.05. 

In the next step, we train the pseudo-encoder with the training dataset and test it with the validation data set using the approach discussed in \cite{kiarashinejad2019deep}. To simplify the computation, we only consider one layer for the encoder part of the pseudo-encoder. Figure \ref{fig:fig5_mse}(b) shows the MSE as a function of the dimension in the reduced design space \textit{D}. It is clear that by reducing the dimension of the design space from 7 to 4, MSE < 0.02 is achieved. 

\begin{figure}
     \centering
    \subfigure[]{\includegraphics[width=0.45\textwidth]{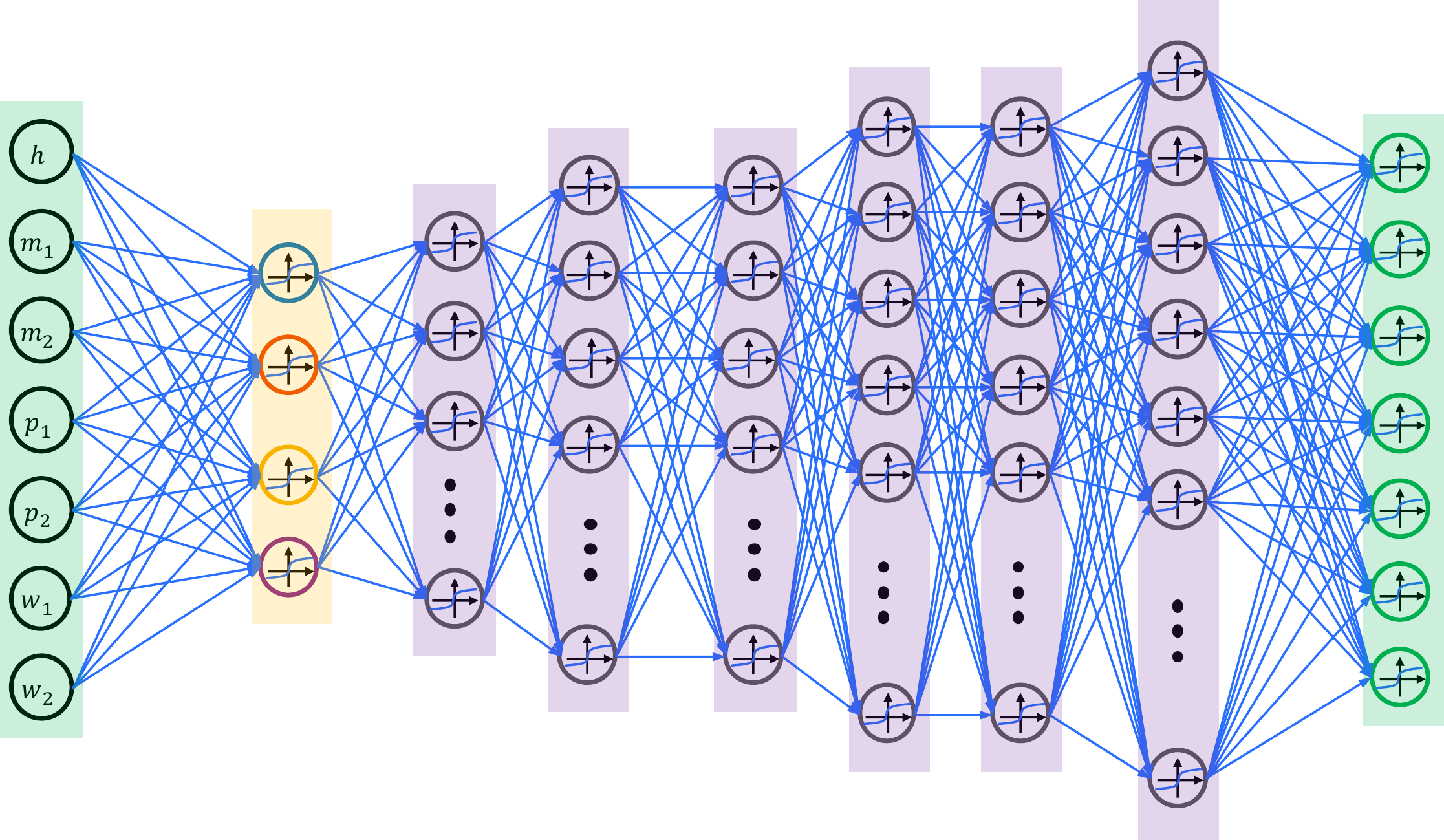}
    \label{fig:Fig.7a}}
    \subfigure[]{\includegraphics[width=0.53\textwidth, trim={1.5cm 0.3cm 2.5cm 0.8cm},clip]{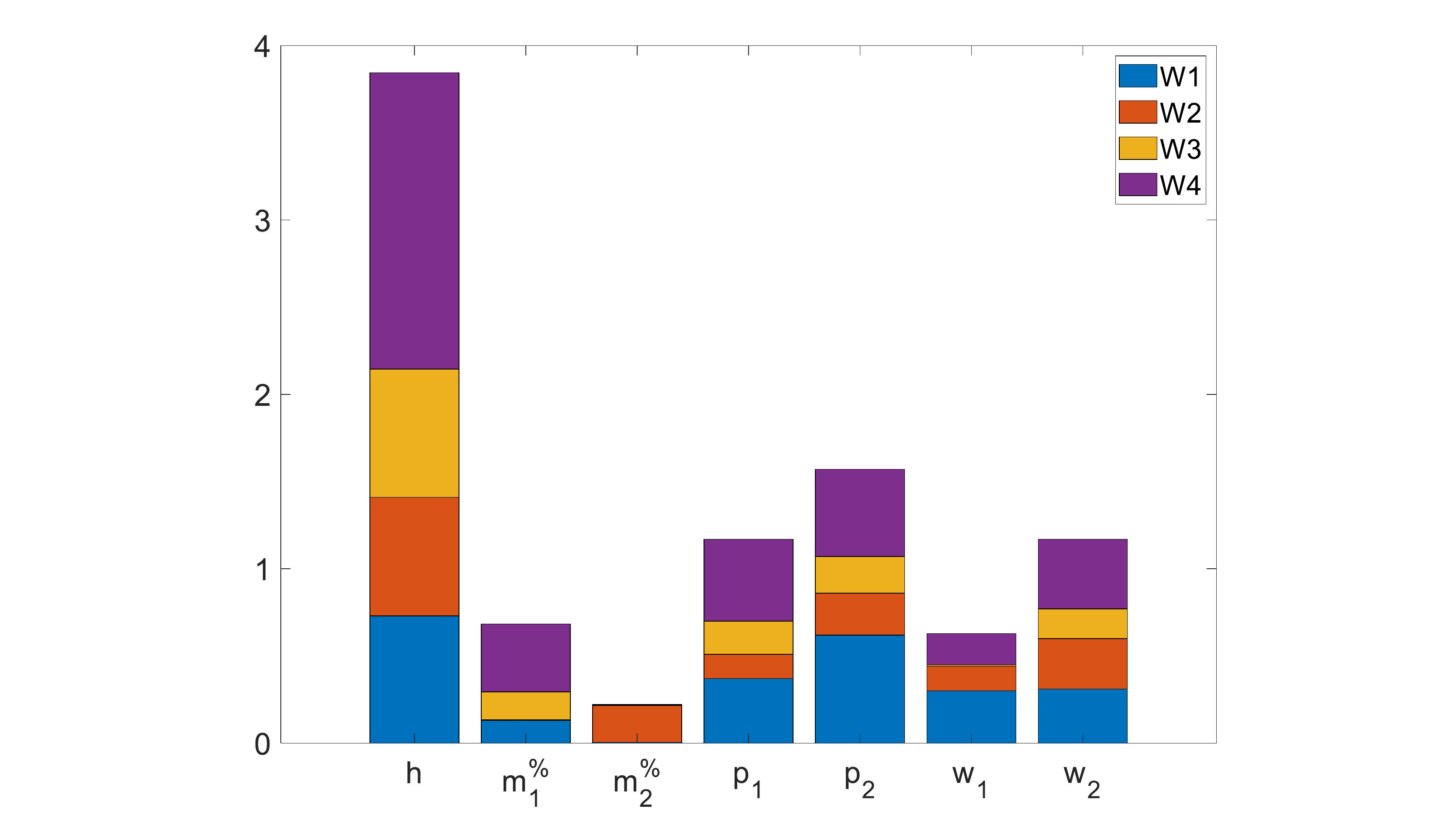}
    \label{fig:Fig.7b}}
  \caption{Detailed architecture of the adopted pseudo-encoder.  (a) The number of nodes of the  pseudo-encoder are 7, 4, 10, 20, 20, 30, 30, 40, and 7 at each layer. (b) Weights of the first layer of the pseudo-encoder in (a) which is yellow highlighted.}
    \label{fig:fig7_weight}
\end{figure}

Figure \ref{fig:fig7_weight}(a) shows the pseudo-encoder architecture with the design parameters and the reduced response space being its input and output, respectively. The weights of the first layer in pseudo-encoder represent the importance of different design parameters. In this manner, each input node (corresponding to each deign parameter) is connected to the nodes in the second layer with the strengths shown by the weights. Thus, design parameters with more significant roles have larger weights in this layer. Figure \ref{fig:fig7_weight}(b) shows the weights in the first layer of the pseudo-encoder. It is clear that the GST thickness $h$ has the strongest role in the output response. Thus, the response of the structure is more sensitive to this parameter. Besides, $w_i$ ($i=1, 2$) and $p_i$ ($i=1, 2$) have almost similar accumulative intensities (or weights) and thus, similar  influence on the response space. It is clear that the response space is slightly affected by the less important design parameter $m_i^{\%}$ ($i=1, 2$). This understanding of the relative importance of the design parameters in the output response is very helpful for initializing any wise optimization process, even with traditional approaches.

\section{Understanding the Physics of Light-matter Interaction in Nanostructures} \label{Understanding}

Herein, we illustrate that the interpretation of the weights of the pseudo-encoder can effectively reveal the underlying physics of light-matter interactions in nanostructures. For this purpose, we perform a comprehensive analysis of the fundamental modes of the metasurface in Fig. \ref{fig:fig4_supercell} using full-wave EM simulations in the given frequency range. Further information on the material properties as well as details of the FEM simulation process are provided in the Methods.

\begin{figure}[h]
\centering
    \subfigure[]{\includegraphics[width=0.3\textwidth, trim={3.5cm 2cm 4cm 2cm},clip]{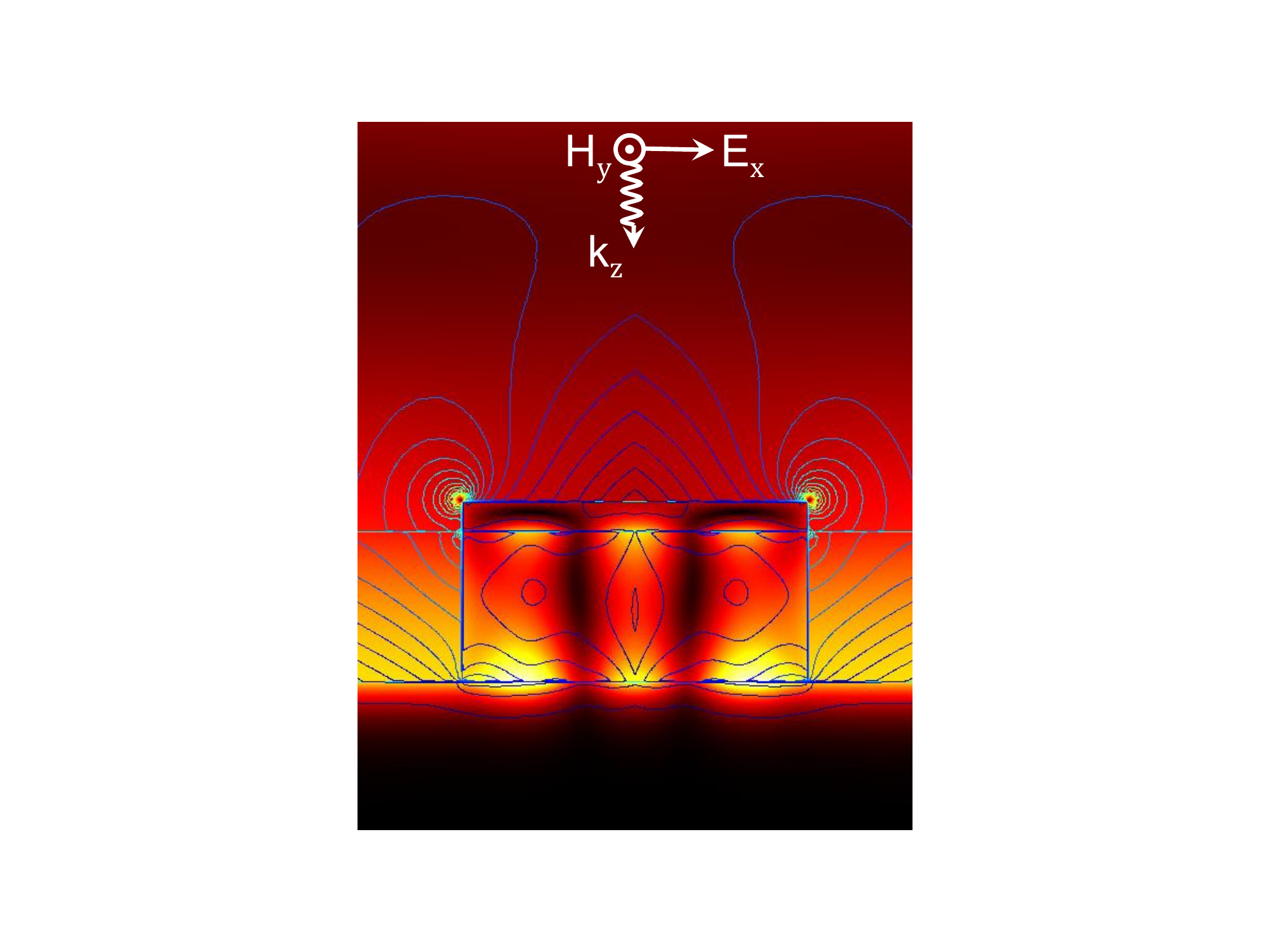}
     \label{fig:Fig.8a}}
    \subfigure[]{\includegraphics[width=0.3\textwidth, trim={3.5cm 2cm 4cm 2cm},clip]{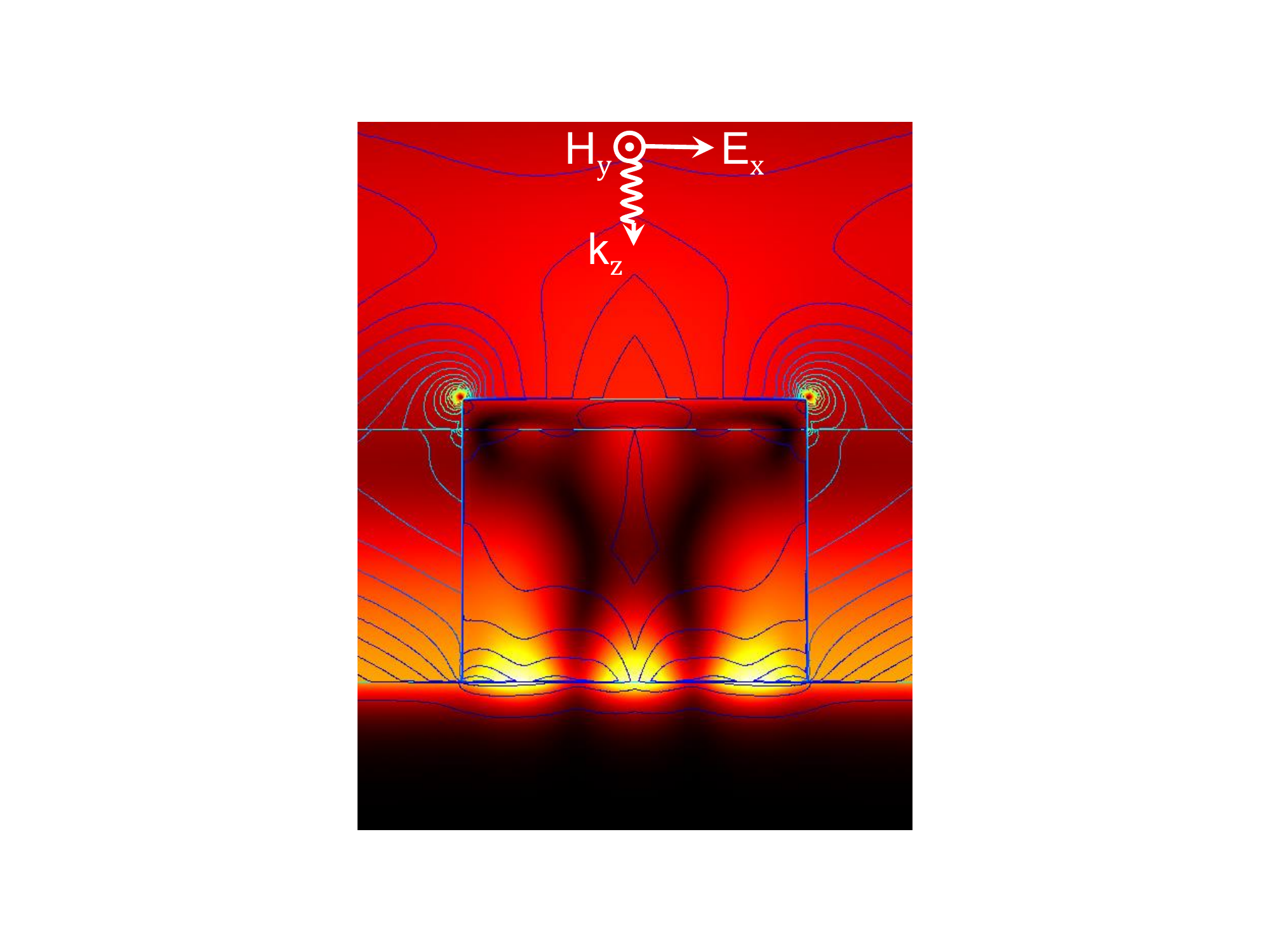}
    \label{fig:Fig.8b}}
    \subfigure[]{\includegraphics[width=0.3\textwidth, trim={3.5cm 2cm 4cm 2cm},clip]{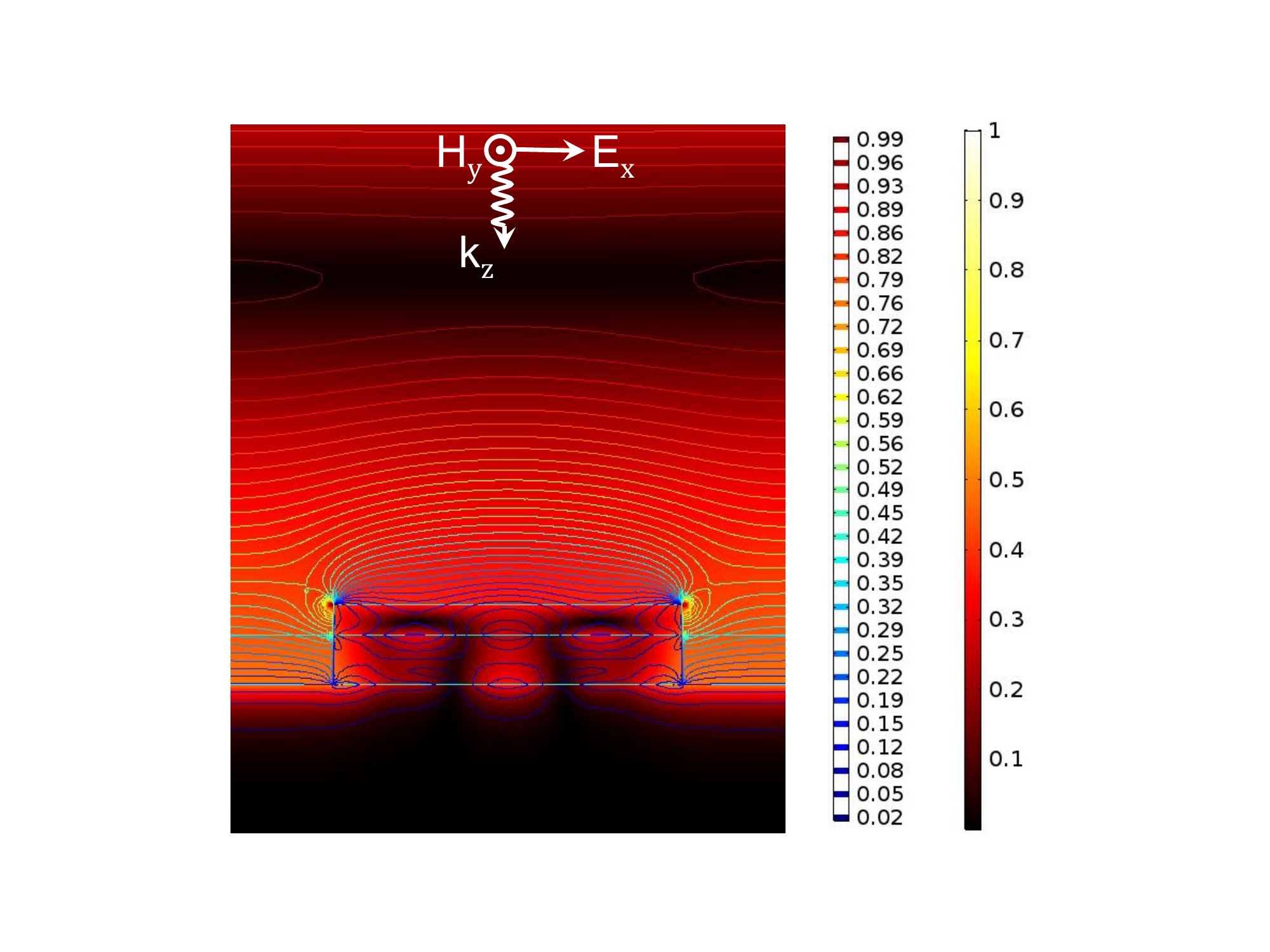}
    \hspace{-1.6cm}
    \label{fig:Fig.8c}}
    \caption{EM field distributions in a unit cell of the structure in Fig. \ref{fig:fig4_supercell}. Magnetic field (presented by the thermal colormap) and electric field profile (represented by arrows and coded by the rainbow colorbar) for a unit cell with (a) $h$ = 150 nm, (b) $h$ = 250 nm, and (c) $h$ = 50 nm, respectively. The other structural parameters are fixed as $p$ = 550 nm, $w$ = 340 nm, and $m^{60\%}$. The frequency of incident TM-polarized light is $f$ = 194 THz.}
    \label{fig:fig8_magfld}
\end{figure}

Figure \ref{fig:fig8_magfld}(a) shows upon excitation of a unit cell of the structure in Fig. \ref{fig:fig4_supercell} with a TM-polarized light, the Au nanoribbon/GST nanostripe and GST nanostripe/Au back-reflector can support short-range surface plasmons (SR-SPPs), which are EM waves bound to and travel along the metal-dielectric interface with short propagation lengths. It should be noted that the effective index contrast at the interfaces of Au nanoribbon end-faces and air implies that each nanoribbon approximately acts as a \textit{lateral} mirror-like cavity. Accordingly, a constructive interference happens between SR-SPP modes travelling  back and forth (i.e., in the x-direction in Fig. \ref{fig:fig4_supercell}) between the two ends of the Au nanoribbon. Furthermore, the difference between the refractive indices of GST and SiO$_2$ enhances the effect of the lateral Fabry-Perot cavity in the intermediate GST nanostripe at the interface between the GST and the (lower) Au back-reflector plane. Thus, a similar mode profile exists within that region, which can be ascribed to a confined constructive SPP. 

The two aforementioned SPPs are the fundamental modes of the metasurface defining any arbitrary response from the structure. We expect that the parameter simultaneously modifying the field profiles of these modes plays the key role in engineering the spectral response of the metasurface. The DR algorithm introduces $h$ as the most influential parameter (see Fig. \ref{fig:fig7_weight}(b)). To justify such a claim, we study the effect of three different $h$ values on the field profile of the metasurface for a fixed set of other parameters. Figure \ref{fig:fig8_magfld}(b) shows that when the two Au-GST interfaces (at the top and bottom of the GST nanostripe) are far (i.e., large  $h$), the fundamental modes are spatially separated in a unit cell. By decreasing the distance of these interfaces (i.e., $h$), the coupling between the SR-SPP and the confined SPP modes sustained by individual interfaces increases until these highly coupled modes form a supermode as the dominant mode of the structure (see Fig. \ref{fig:fig8_magfld}(a)). Fig. \ref{fig:fig8_magfld}(c) shows that further decrease in parameter $h$ results in fading of one of the modes. To further verify our conclusion, we finely sweep $h$ and present the reflection spectrum in Fig. \ref{fig:Fig.10a}. This figure illustrates that by changing $h$, no abrupt change happens in the reflection spectrum profile. Such a gradual transition verifies the absence of the well-known gap-surface-plasmon resonance, a highly confined magnetic mode, which originates from the circulating displacement current between a metal nanoribbon and a metal back-reflector. This can be ascribed to the remarkable refractive index of GST in any crystallization fraction. Thus, we firmly conclude the metasurface only supports the two above-mentioned SPP modes (and no gap-surface-plasmon mode). This makes $h$ the most important design parameter with the maximum influence on the variation of the reflection spectrum.

\begin{figure}[h]
\centering
    \subfigure[]{\includegraphics[width=0.3\textwidth, trim={3.5cm 2cm 4cm 2cm},clip]{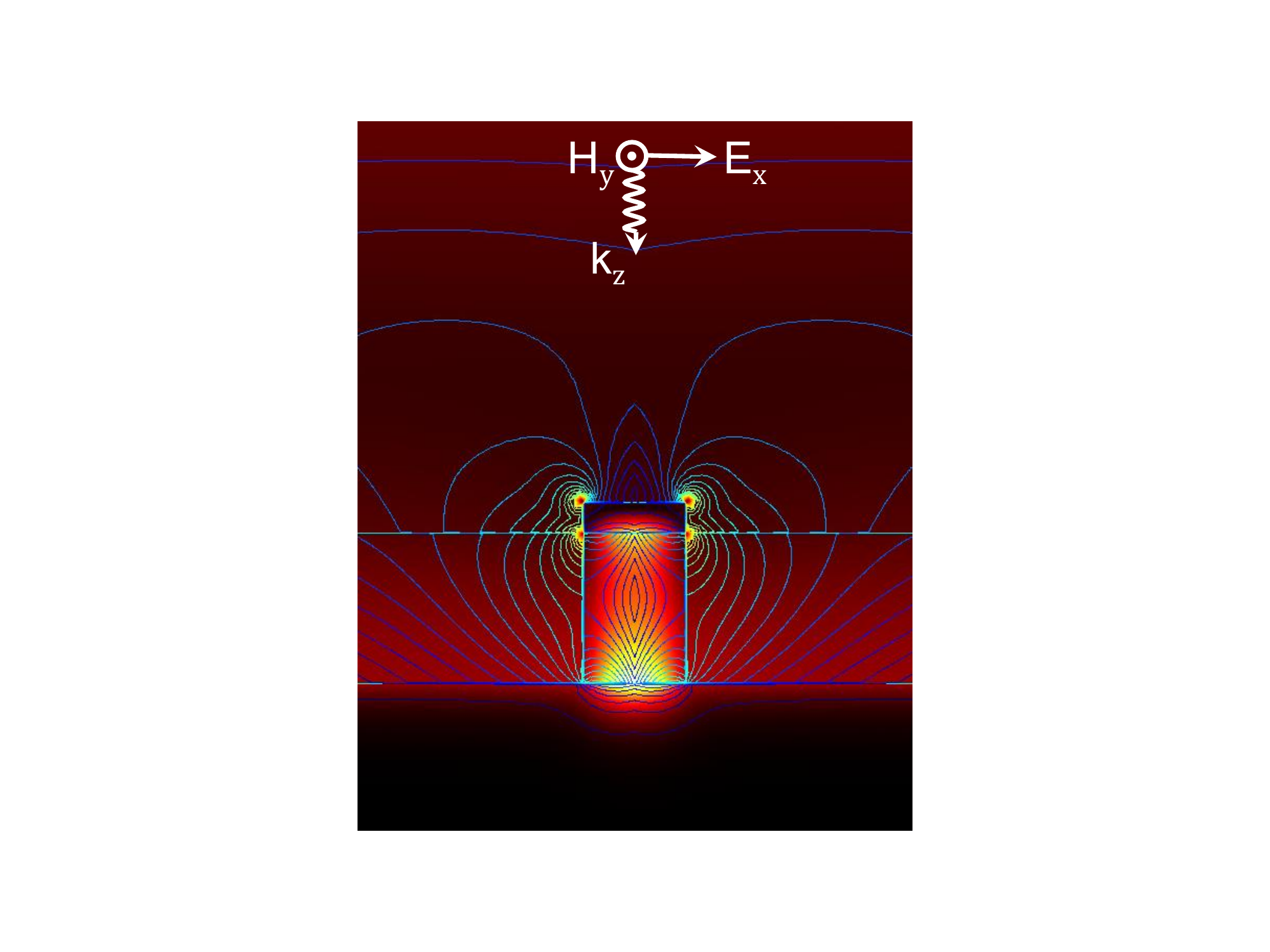}
     \label{fig:Fig.9a}}
    \subfigure[]{\includegraphics[width=0.3\textwidth, trim={3.5cm 2cm 4cm 2cm},clip]{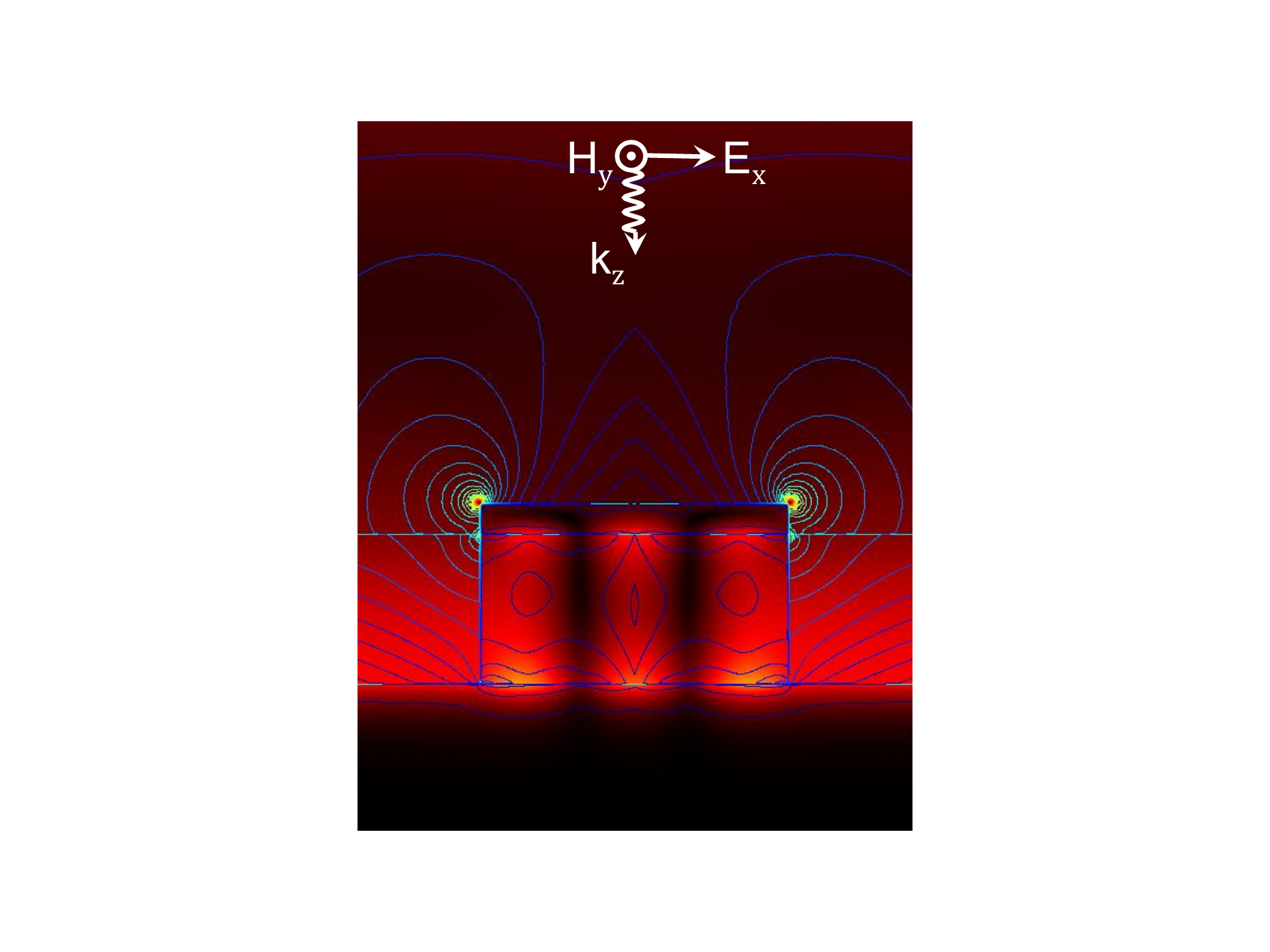}
    \label{fig:Fig.9b}}
    \subfigure[]{\includegraphics[width=0.3\textwidth, trim={3.5cm 2cm 4cm 2cm},clip]{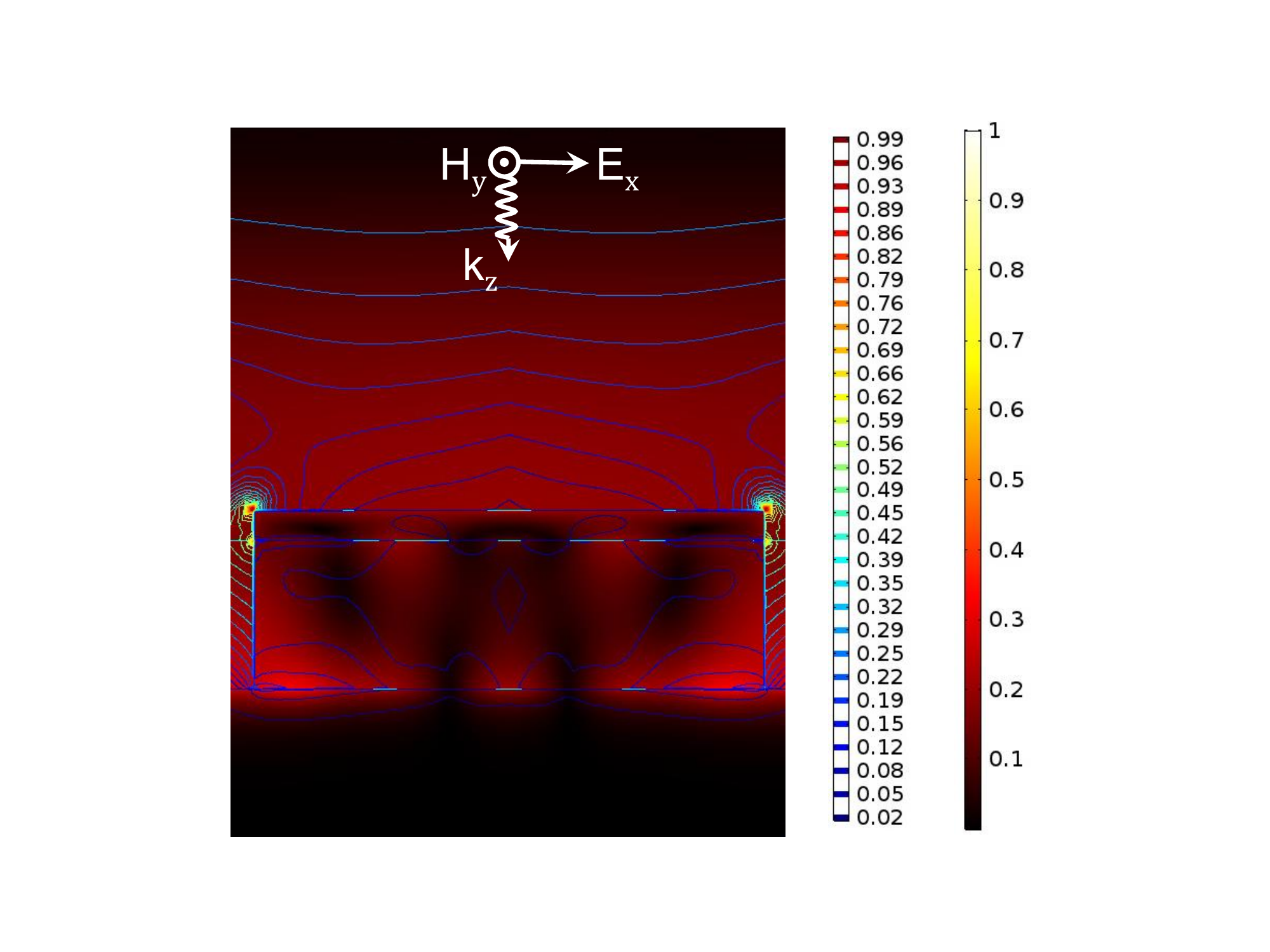}
    \hspace{-1.6cm}
    \label{fig:Fig.9c}}
    \caption{EM field distributions in a unit cell of the structure in Fig. \ref{fig:fig4_supercell}. Magnetic field (presented by the thermal colormap) and electric field profile (represented by arrows and coded by the rainbow colorbar) for a unit cell with (a) $w$ = 100 nm (first order mode), (b) $w$ = 300 nm (third order mode), and (c) $w$ = 500 nm (fifth order mode), respectively. The other structural parameters are fixed as $p$ = 550 nm, $h$ = 150 nm, and $m^{60\%}$. The frequency of incident TM-polarized light is $f$ = 194 THz.}
    \label{fig:fig9}
\end{figure}

\begin{figure}
    \centering
    \subfigure[]{\includegraphics[width=0.45\textwidth]{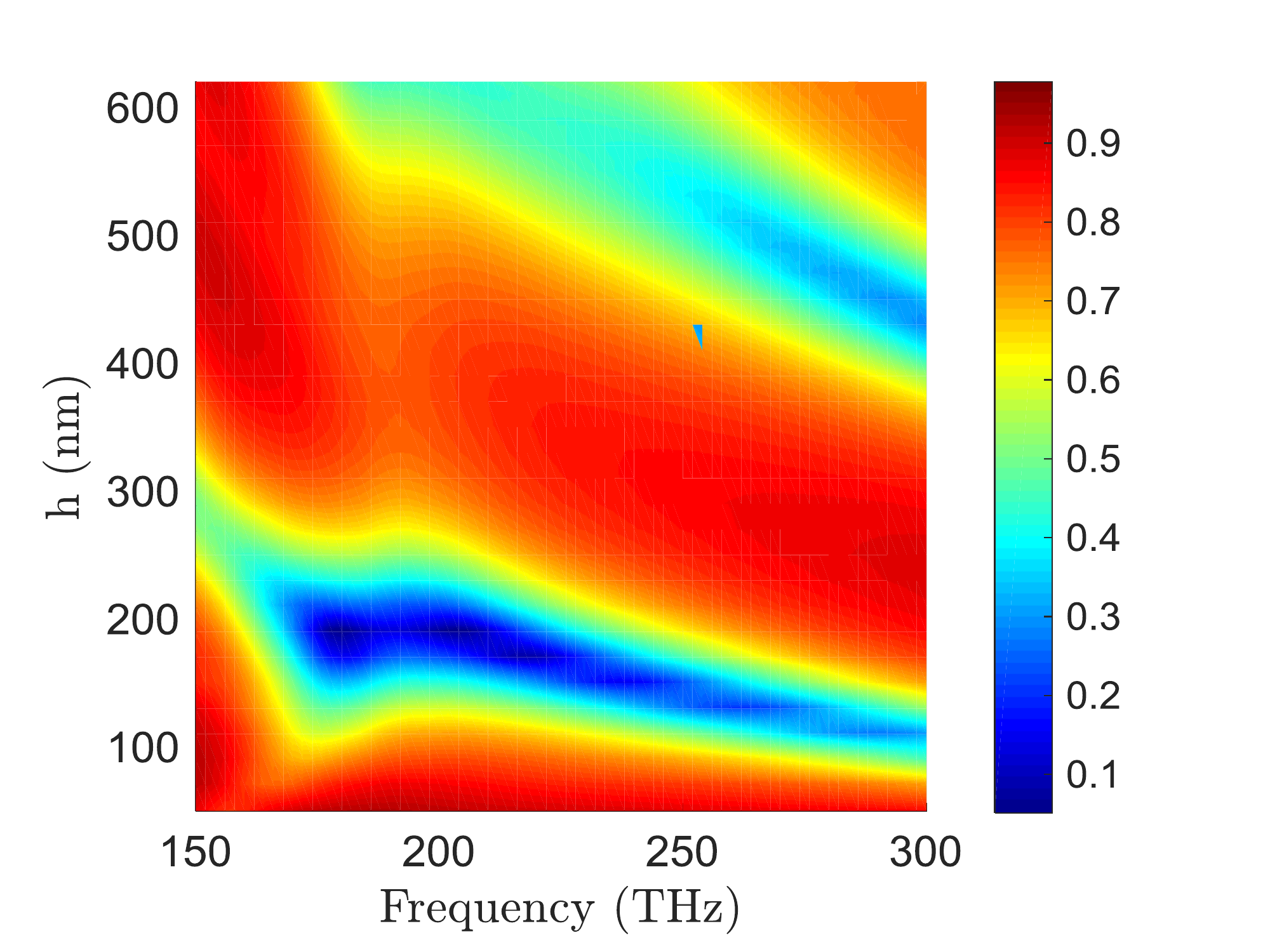}
    \label{fig:Fig.10a}}
    \subfigure[]{\includegraphics[width=0.45\textwidth]{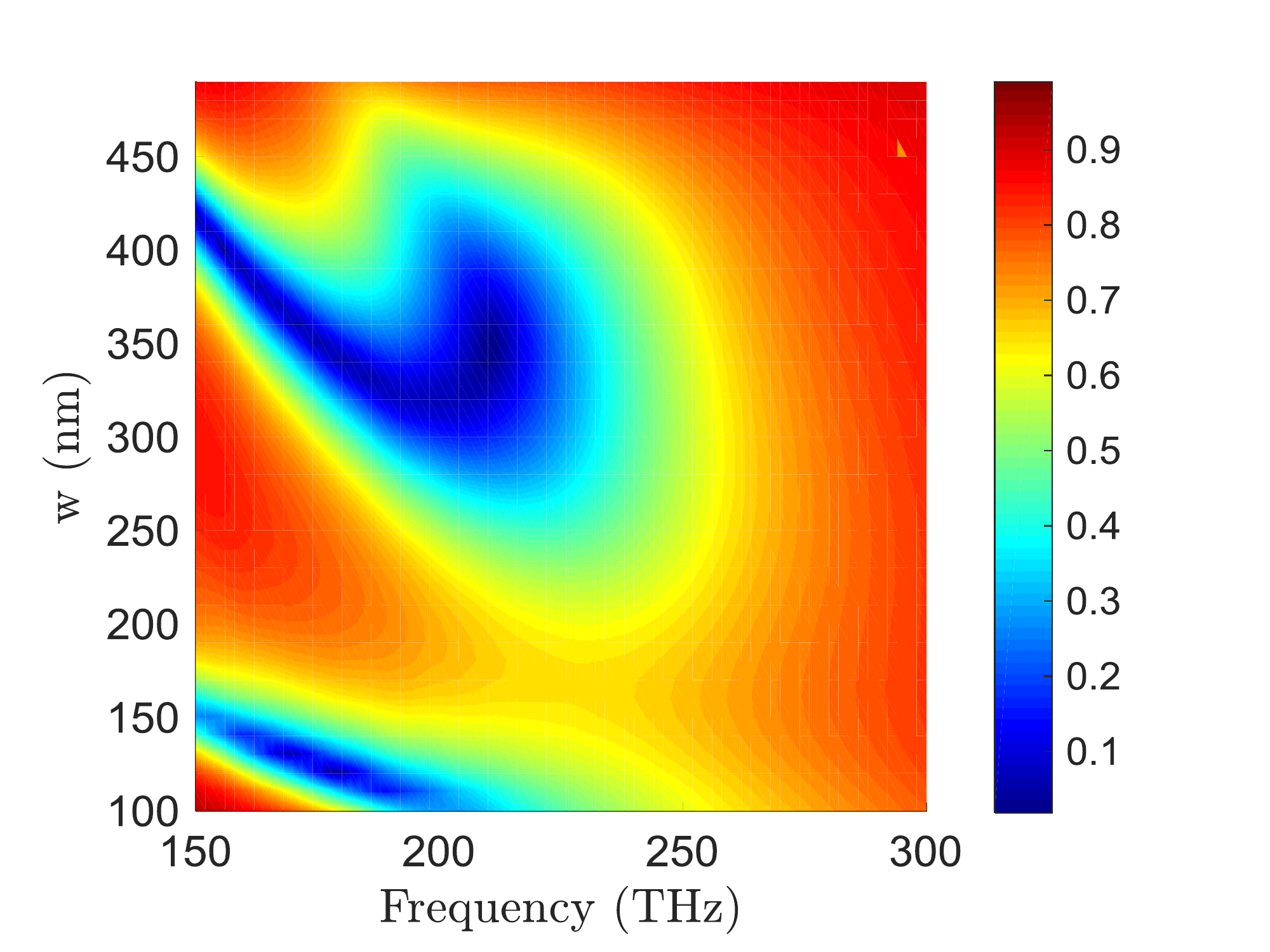}
    \label{fig:Fig.10b}}
    \subfigure[]{\includegraphics[width=0.45\textwidth]{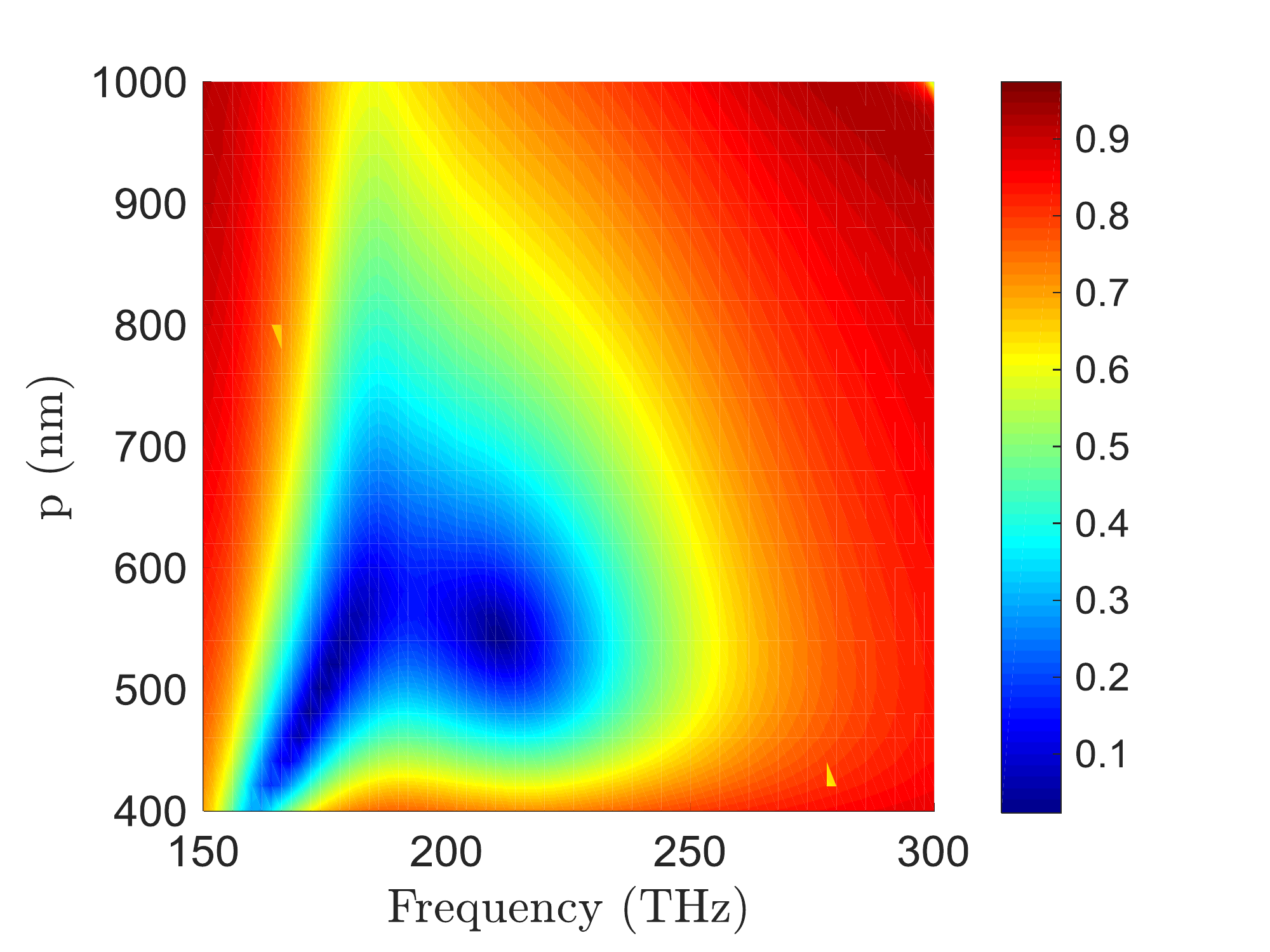}
    \label{fig:Fig.10c}}
    \subfigure[]{\includegraphics[width=0.45\textwidth]{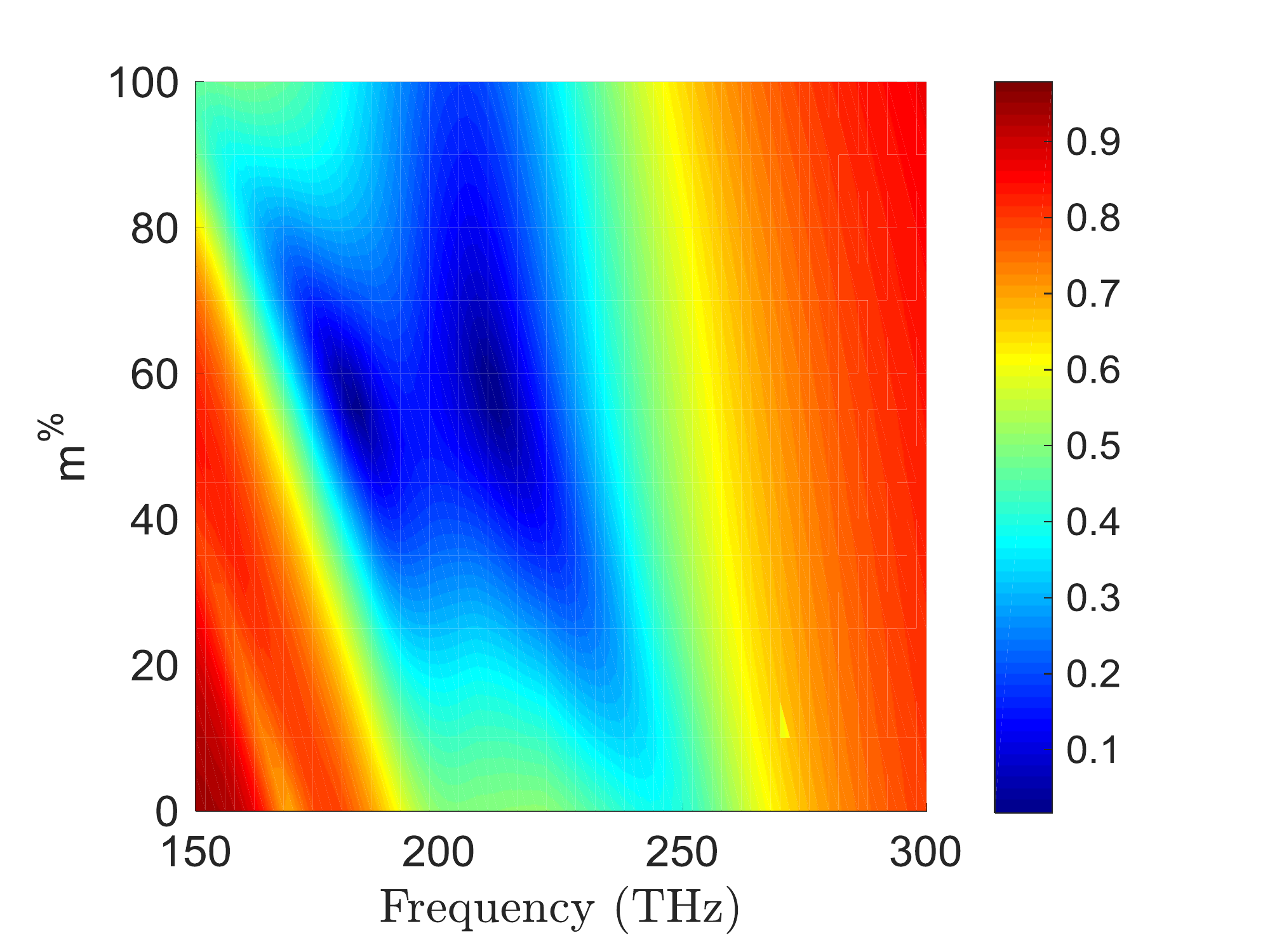}
    \label{fig:Fig.10d}}
    \caption[short]{Reflection amplitude profile of a unit cell of the metasurface shown in Fig. \ref{fig:fig4_supercell} with different structural parameters under illumination of a TM-polarized light. Reflection amplitude versus (a) GST nanostripe thicknesses while geometrical parameters are chosen $p$ = 580 nm, $w$ = 350 nm, and $m^{0\%}$, (b) Au nanoribbon widths while other geometrical parameters are fixed as $h$ = 170 nm, $p$ = 580 nm, and $m^{0\%}$, (c) periodicity of the unit cell while other geometrical parameters are chosen $h$ = 170 nm, $w$ = 350 nm, and $m^{0\%}$, and (d) crystallization fraction while geometrical parameters are fixed as $h$ = 170 nm, $p$ = 580 nm, and $w$ = 350 nm.}
    \label{fig:Fig.10}
\end{figure}

Figure \ref{fig:fig7_weight}(b), shows that $w_i$ ($i=1, 2$) and $p_i$ ($i=1, 2$) have the secondary dominant effects on the EM response after $h$. This is in-line with the well-known fact that the resonance frequency of a SPP mode is highly dependent on the width of the nanocavity. Moreover, we found that even by continuously increasing $w$, only the well-known odd order SPP modes could be excited (see Fig. \ref{fig:fig9}) \cite{della2008plasmon}. This observation justifies that variation of $w$ does not change the inherent nature of these individual SPP modes and keeps them rather decoupled. Figure \ref{fig:Fig.10b} corroborates that while other parameters in a unit cell are fixed, decreasing (increasing) the width ($w$) reasonably blueshifts (redshifts) the resonance. However, the reflection response is not as sensitive to $w$ as to $h$ since the nature of both individual SPP modes remains intact while $w$ is varied. More apparently, by modifying the width around $w$ = 350 nm while having other parameters fixed (i.e., $h$ = 170 nm, $p$ = 580 nm, and $m^{0\%}$), the resonance profile in Fig. \ref{fig:Fig.10b} has a broader linewidth compared to its counterpart in Fig. \ref{fig:Fig.10a}. This comparison well justifies the more sensitive nature of the reflection spectrum to $h$. On the role of $p$, it is notable that light diffraction from the surface of the structure is the origination of the confined SPP mode excited at the interface of the GST nanostripe/Au back-reflector. As a result, the behavior of the overall reflection spectrum relies on the periodicity (i.e., $p$) of each unit cell. Figure \ref{fig:Fig.10c} illustrates that by increasing $p$, the lower portion of the incident light couples to the confined SPP mode, and some part of it reflects in the form of higher diffraction orders. On the other hand, decreasing the periodicity can reasonably increase the coupling of adjacent unit cells, which changes the overall reflection response. 

Finally, Fig. \ref{fig:fig7_weight}(b) shows that $m_i^{\%}$ ($i=1, 2$) has the minimum effect on the EM response among the investigated parameters. This is also seen from Fig. \ref{fig:Fig.10d} as slightly changing $m$ (while keeping all other parameters fixed) results in no major change in the reflection spectrum amplitude. Nevertheless, Fig. \ref{fig:Fig.10d} suggests that the location of the minimum of the reflection spectrum (or the absorption peak) depends on $m^{\%}$; increasing (decreasing) $m^{\%}$ (i.e., larger GST refractive index) results in a red (blue) shift in the absorption peak. Accordingly, choosing a unit cell with a proper crystallization fraction ($m^{\%}$) ensures near-unity absorption at the desired frequency. More importantly, a structure with a supercell with two different values of $m$ ($m_1^{\%}$ and $m_2^{\%}$ in Fig. \ref{fig:fig4_supercell}) can have a multi-band absorption governed by the constructive and/or destructive overlap between the distinct resonance peaks corresponding to the two values of $m^{\%}$. This design approach can be extended to structures with more sophisticated supercells to form more complex absorption spectra.

\begin{figure}
\centering
    \subfigure[]{\includegraphics[width=0.45\textwidth]{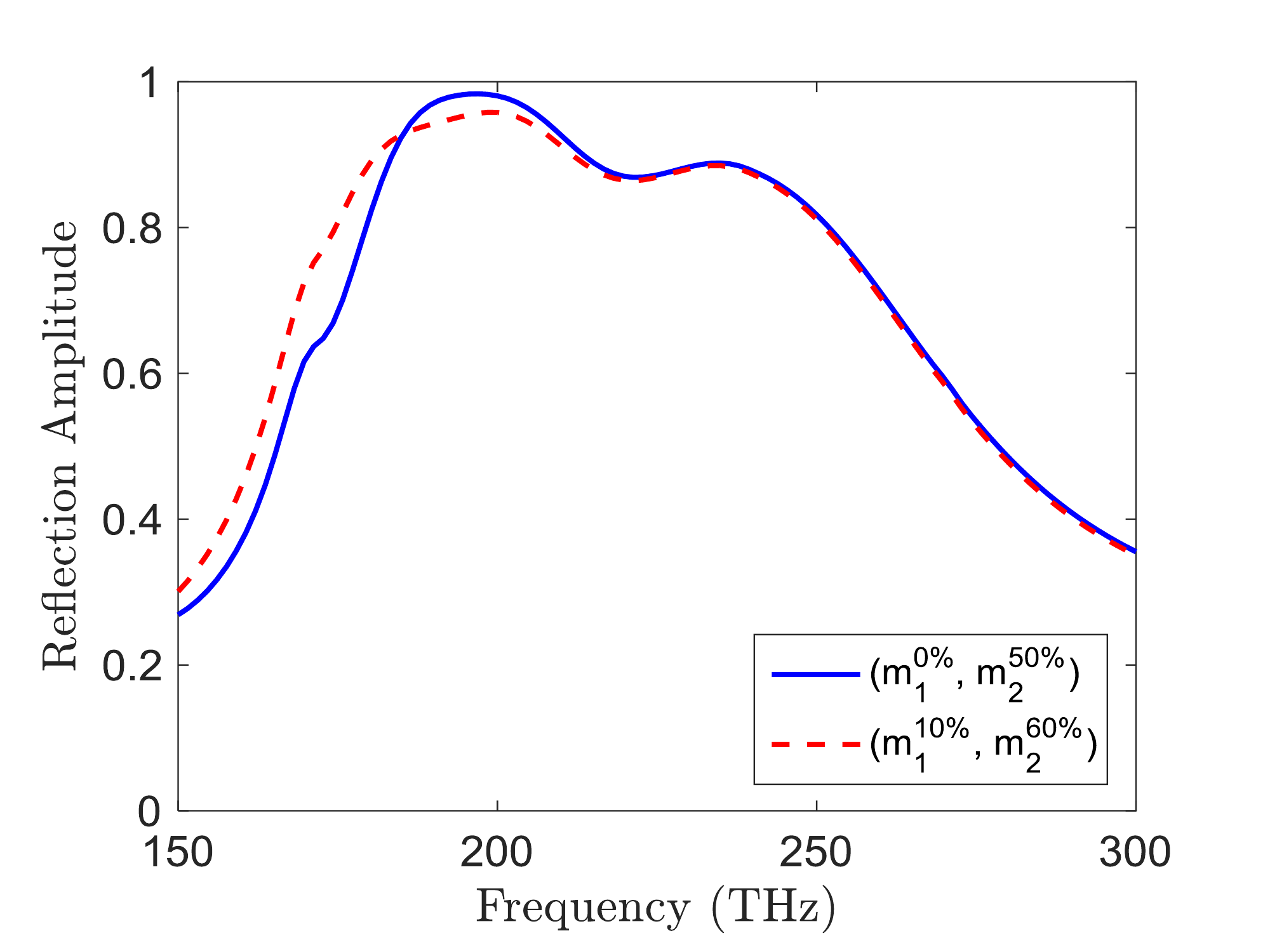}
     \label{fig:Fig.11a}}
    \subfigure[]{\includegraphics[width=0.45\textwidth]{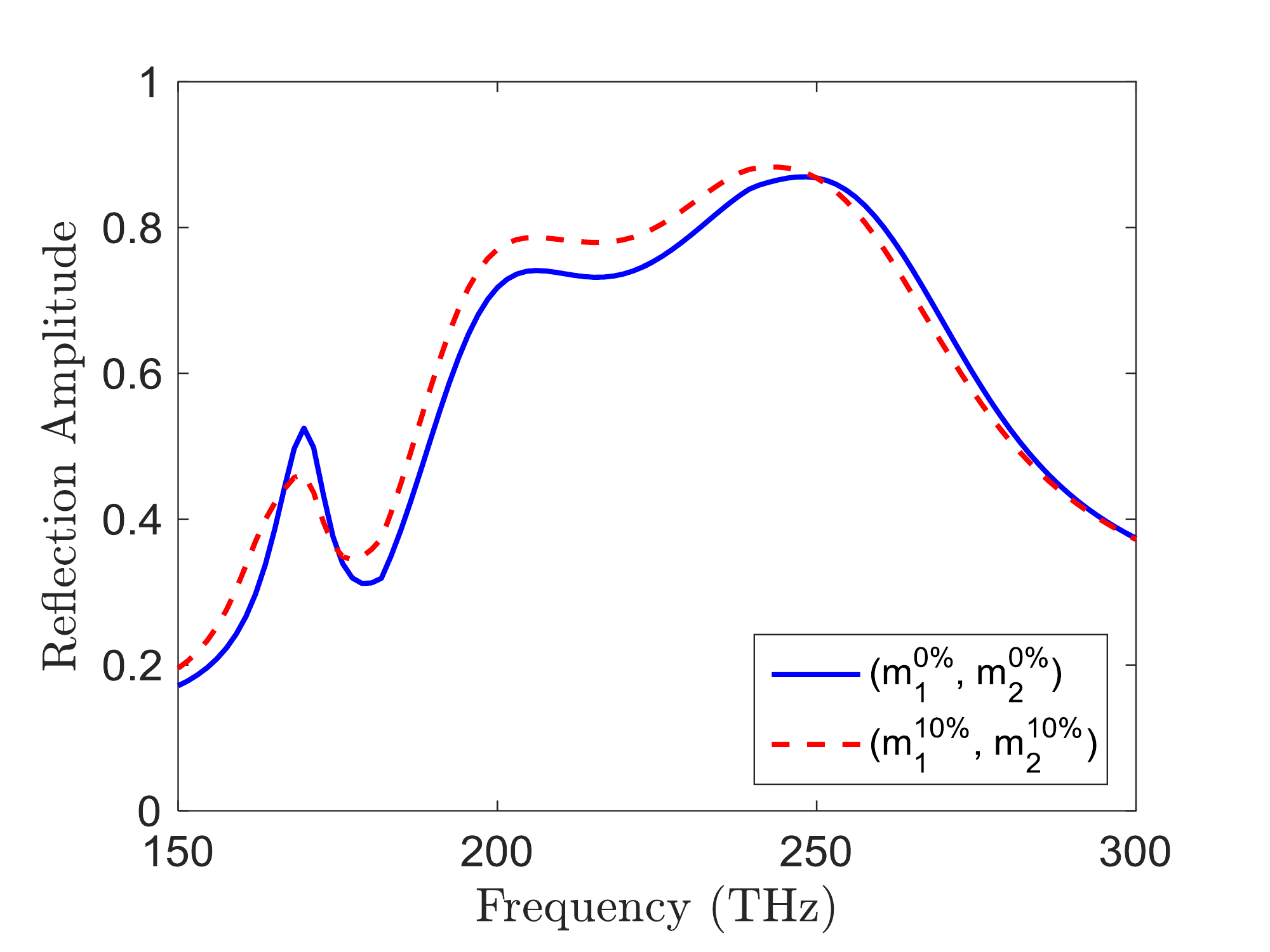}
    \label{fig:Fig.11b}}
    \caption{Absorption spectra of the designed optimized multifunctional metadevice. (a) Spectra of the original dual-band absorber (solid blue line) at two distinct frequencies of $f_1$ = 195 THz and $f_2$ =235 THz, and (b) triple-band absorber (solid blue line) at three distinct frequencies of $f_1$ = 170 THz, $f_2$ = 205 THz, and $f_3$ = 240 THz. Red dashed curves justify that by slightly modifying $m_i^{\%}$ ($i=1, 2$) around the optimized values, the spectra change moderately.}
    \label{fig:fig11_Rflc}
\end{figure}

\section{Discussion} \label{Discus}

The intuitive understanding of the role of design parameters on the response of a given nanostructure obtained in Section \ref{Understanding} can facilitate the design of more sophisticate nanostructures using any optimization technique. The computation complexity of any optimization approach depends heavily on the discretization of the values of different design parameters. Our approach suggests that maximum discretization shall be used for the most influential design parameter (e.g., $h$ in the structure in Fig. \ref{fig:fig4_supercell}) while a more sparse discretization is acceptable for less important design parameters (e.g., $m_i^{\%}$ ($i=1, 2$) in the structure in Fig. \ref{fig:fig4_supercell}). To test this approach, we used our findings in Section \ref{Understanding} about the structure in Fig. \ref{fig:fig4_supercell} to design a multifunctional metadevice providing dual-band absorption (at two distinct frequencies of $f_1$ = 195 THz and $f_2$ =235 THz) and triple-band absorption (at three distinct frequencies of $f_1$ = 170 THz, $f_2$ = 205 THz, and $f_3$ = 240 THz). For device optimization, we used the exhaustive search of design parameters using the analytic formulas obtained by the trained system that relates the design space to the response space with DR. We also used non-uniform discretization of the design parameters through our findings in Section \ref{Understanding} to minimize the computation complexity. The optimized supercell offered by this approach has two unit cells with similar structural parameters $h$ = 180 nm, $p_i$ = 550 nm ($i=1, 2$), and $w_i$ = 340 nm ($i=1, 2$).  ($m_{1}^{0\%}$,$m_{2}^{50\%}$) and ($m_{1}^{0\%}$,$m_{2}^{0\%}$) are obtained for the dual- and the triple-band absorption functionalities, respectively. Figure \ref{fig:fig11_Rflc} shows the response of the designed multifunctional metadevice. As shown, by slightly changing $m_{1}^{\%}$ and $m_{2}^{\%}$, the locations of the absorption peaks do not considerably change around the optimized values. This is anticipated regarding the discussion in Section \ref{Understanding};  the overall reflection response of the structure has minimum sensitivity to $m_{1}^{\%}$ and $m_{2}^{\%}$ and thus, it is robust against random variations of external stimulus (here the gate voltage) or other destructive environmental effects (such as GST oxidization).

Another important observation is that by retraining the algorithm using a different set of training data (for the same structure), the weights slightly change, but the trends of their variations remain the same.  This means that the intuitive understanding of the roles of the designing parameters in its response is to a good degree independent of the training process. We repeated this process at least 20 times with different sets of training data for the structure in Fig. \ref{fig:fig4_supercell} and found the same conclusions in all trials. Nevertheless, we think that for sophisticated nanostructures with many design parameters, training the algorithm with different training sets may reveal different information about the device operation. This is especially valuable for the complex nanostructures in which simple simulations (e.g., like the ones that resulted in Fig. \ref{fig:fig11_Rflc}) cannot be used to understand the role of design parameters. It is also not practical to simulate enough structures to learn the role of a specific design parameter due to computation complexity of such complex nanostructures. It is also worth mentioning that the technique discussed here is not limited to nanostructures; it can be extended to cover many different problems (e.g., fluid mechanics, heat transfer, acoustic wave propagation, etc.) as long as enough training data can be provided.

\section{Conclusion} \label{Conclusion}

We demonstrated here a DL-based technique for the understanding of the physics of wave-matter interaction in nanostructures. By using the DR algorithm in the response space and the design space (using an autoencoder and a pseudo-encoder, respectively), we could obtain an analytic formula that relates the design parameters to the response of the nanostructure while providing access to the weights of the neural notworks at all layers. By analyzing these weights, important information about the roles of different design parameters in the overall response of the nanostructure can be obtained. This intuitive information can be used to understand the physics of light-matter interaction while facilitating the device optimization process by suggesting a non-uniform discretization of the design parameters to reduce the computation requirements. As such, the approach presented here can have an important impact in the design and understanding of the EM wave-matter interaction in nanostructures while being extendable to several other applications. 

\section*{Methods}

All full-wave EM simulation results shown in the main text were obtained by using COMSOL Multiphysics 5.3, a commercialized full-wave simulation package based on the FEM. The proposed metasurface in Fig. \ref{fig:fig4_supercell} was simulated in a two-dimensional environment with periodic boundary conditions in the y direction. The structure is assumed infinite in the x direction and was excited with a TM-polarized plane-wave propagating in the +z direction. The optical properties of the amorphous and the fully crystalline GST were obtained from \cite{shportko2008resonant} and those of the intermediate states were calculated using the well-known Lorentz-Lorenz relation formulated as \cite {abdollahramezani2018reconfigurable}: 

\begin{equation} 
\frac{\epsilon_{eff}(f)-1}{\epsilon_{eff}(f)+2}={{(m^{\%}/100)}\times\frac{\epsilon_{c}(f)-1}{\epsilon_{c}(f)+2}+({(m^{\%}/100)}-1)\times\frac{\epsilon_{a}(f)-1}{\epsilon_{a}(f)+2}},
\label{Lorentz-Lorenz}
\end{equation}
where for a specific frequency $f$, $\epsilon_{c}(f)$ and $\epsilon_{a}(f)$ are the permittivities of the fully crystalline and the amorphous GST, respectively, and ${m^{\%}}$, ranging from ${0\%}$ (i.e., ${m^{0\%}}$ or amorphous) to ${100\%}$ (i.e., ${m^{100\%}}$, fully crystalline), is the crystallization fraction of GST. The optical properties of other materials were obtained from \cite{yu2011light}.



\bibliography{sample}

\begin{thebibliography}{10}
	\urlstyle{rm}
	\expandafter\ifx\csname url\endcsname\relax
	\def\url#1{\texttt{#1}}\fi
	\expandafter\ifx\csname urlprefix\endcsname\relax\def\urlprefix{URL }\fi
	\expandafter\ifx\csname doiprefix\endcsname\relax\def\doiprefix{DOI: }\fi
	\providecommand{\bibinfo}[2]{#2}
	\providecommand{\eprint}[2][]{\url{#2}}
	
	\bibitem{jahani2016all}
	\bibinfo{author}{Jahani, S.} \& \bibinfo{author}{Jacob, Z.}
	\newblock \bibinfo{journal}{\bibinfo{title}{All-dielectric metamaterials}}.
	\newblock {\emph{\JournalTitle{Nature nanotechnology}}}
	\textbf{\bibinfo{volume}{11}}, \bibinfo{pages}{23} (\bibinfo{year}{2016}).
	
	\bibitem{yu2011light}
	\bibinfo{author}{Yu, N.} \emph{et~al.}
	\newblock \bibinfo{journal}{\bibinfo{title}{Light propagation with phase
			discontinuities: generalized laws of reflection and refraction}}.
	\newblock {\emph{\JournalTitle{science}}} \bibinfo{pages}{1210713}
	(\bibinfo{year}{2011}).
	
	\bibitem{arbabi2015dielectric}
	\bibinfo{author}{Arbabi, A.}, \bibinfo{author}{Horie, Y.},
	\bibinfo{author}{Bagheri, M.} \& \bibinfo{author}{Faraon, A.}
	\newblock \bibinfo{journal}{\bibinfo{title}{Dielectric metasurfaces for
			complete control of phase and polarization with subwavelength spatial
			resolution and high transmission}}.
	\newblock {\emph{\JournalTitle{Nature nanotechnology}}}
	\textbf{\bibinfo{volume}{10}}, \bibinfo{pages}{937} (\bibinfo{year}{2015}).
	
	\bibitem{hsiao2017fundamentals}
	\bibinfo{author}{Hsiao, H.-H.}, \bibinfo{author}{Chu, C.~H.} \&
	\bibinfo{author}{Tsai, D.~P.}
	\newblock \bibinfo{journal}{\bibinfo{title}{Fundamentals and applications of
			metasurfaces}}.
	\newblock {\emph{\JournalTitle{Small Methods}}} \textbf{\bibinfo{volume}{1}},
	\bibinfo{pages}{1600064} (\bibinfo{year}{2017}).
	
	\bibitem{khorasaninejad2016metalenses}
	\bibinfo{author}{Khorasaninejad, M.} \emph{et~al.}
	\newblock \bibinfo{journal}{\bibinfo{title}{Metalenses at visible wavelengths:
			Diffraction-limited focusing and subwavelength resolution imaging}}.
	\newblock {\emph{\JournalTitle{Science}}} \textbf{\bibinfo{volume}{352}},
	\bibinfo{pages}{1190--1194} (\bibinfo{year}{2016}).
	
	\bibitem{abdollahramezani2015analog}
	\bibinfo{author}{AbdollahRamezani, S.}, \bibinfo{author}{Arik, K.},
	\bibinfo{author}{Khavasi, A.} \& \bibinfo{author}{Kavehvash, Z.}
	\newblock \bibinfo{journal}{\bibinfo{title}{Analog computing using
			graphene-based metalines}}.
	\newblock {\emph{\JournalTitle{Optics letters}}} \textbf{\bibinfo{volume}{40}},
	\bibinfo{pages}{5239--5242} (\bibinfo{year}{2015}).
	
	\bibitem{chizari2016analog}
	\bibinfo{author}{Chizari, A.}, \bibinfo{author}{Abdollahramezani, S.},
	\bibinfo{author}{Jamali, M.~V.} \& \bibinfo{author}{Salehi, J.~A.}
	\newblock \bibinfo{journal}{\bibinfo{title}{Analog optical computing based on a
			dielectric meta-reflect array}}.
	\newblock {\emph{\JournalTitle{Optics letters}}} \textbf{\bibinfo{volume}{41}},
	\bibinfo{pages}{3451--3454} (\bibinfo{year}{2016}).
	
	\bibitem{abdollahramezani2017dielectric}
	\bibinfo{author}{Abdollahramezani, S.}, \bibinfo{author}{Chizari, A.},
	\bibinfo{author}{Dorche, A.~E.}, \bibinfo{author}{Jamali, M.~V.} \&
	\bibinfo{author}{Salehi, J.~A.}
	\newblock \bibinfo{journal}{\bibinfo{title}{Dielectric metasurfaces solve
			differential and integro-differential equations}}.
	\newblock {\emph{\JournalTitle{Optics letters}}} \textbf{\bibinfo{volume}{42}},
	\bibinfo{pages}{1197--1200} (\bibinfo{year}{2017}).
	
	\bibitem{campbell2019review}
	\bibinfo{author}{Campbell, S.~D.} \emph{et~al.}
	\newblock \bibinfo{journal}{\bibinfo{title}{Review of numerical optimization
			techniques for meta-device design}}.
	\newblock {\emph{\JournalTitle{Optical Materials Express}}}
	\textbf{\bibinfo{volume}{9}}, \bibinfo{pages}{1842--1863}
	(\bibinfo{year}{2019}).
	
	\bibitem{sakurai2019ultranarrow}
	\bibinfo{author}{Sakurai, A.} \emph{et~al.}
	\newblock \bibinfo{journal}{\bibinfo{title}{Ultranarrow-band
			wavelength-selective thermal emission with aperiodic multilayered
			metamaterials designed by bayesian optimization}}.
	\newblock {\emph{\JournalTitle{ACS central science}}}
	\textbf{\bibinfo{volume}{5}}, \bibinfo{pages}{319--326}
	(\bibinfo{year}{2019}).
	
	\bibitem{pestourie2018inverse}
	\bibinfo{author}{Pestourie, R.} \emph{et~al.}
	\newblock \bibinfo{journal}{\bibinfo{title}{Inverse design of large-area
			metasurfaces}}.
	\newblock {\emph{\JournalTitle{Optics express}}} \textbf{\bibinfo{volume}{26}},
	\bibinfo{pages}{33732--33747} (\bibinfo{year}{2018}).
	
	\bibitem{ma2018deep}
	\bibinfo{author}{Ma, W.}, \bibinfo{author}{Cheng, F.} \& \bibinfo{author}{Liu,
		Y.}
	\newblock \bibinfo{journal}{\bibinfo{title}{Deep-learning-enabled on-demand
			design of chiral metamaterials}}.
	\newblock {\emph{\JournalTitle{ACS nano}}} \textbf{\bibinfo{volume}{12}},
	\bibinfo{pages}{6326--6334} (\bibinfo{year}{2018}).
	
	\bibitem{chen2013high}
	\bibinfo{author}{Chen, W.~T.} \emph{et~al.}
	\newblock \bibinfo{journal}{\bibinfo{title}{High-efficiency broadband
			meta-hologram with polarization-controlled dual images}}.
	\newblock {\emph{\JournalTitle{Nano letters}}} \textbf{\bibinfo{volume}{14}},
	\bibinfo{pages}{225--230} (\bibinfo{year}{2013}).
	
	\bibitem{taghinejad2018ultrafast}
	\bibinfo{author}{Taghinejad, M.} \emph{et~al.}
	\newblock \bibinfo{journal}{\bibinfo{title}{Ultrafast control of phase and
			polarization of light expedited by hot-electron transfer}}.
	\newblock {\emph{\JournalTitle{Nano letters}}} \textbf{\bibinfo{volume}{18}},
	\bibinfo{pages}{5544--5551} (\bibinfo{year}{2018}).
	
	\bibitem{taghinejad2018hot}
	\bibinfo{author}{Taghinejad, M.} \emph{et~al.}
	\newblock \bibinfo{journal}{\bibinfo{title}{Hot-electron-assisted femtosecond
			all-optical modulation in plasmonics}}.
	\newblock {\emph{\JournalTitle{Advanced Materials}}}
	\textbf{\bibinfo{volume}{30}}, \bibinfo{pages}{1704915}
	(\bibinfo{year}{2018}).
	
	\bibitem{Seidel1994}
	\bibinfo{author}{Seidel, S.~Y.} \& \bibinfo{author}{Rappaport, T.~S.}
	\newblock \bibinfo{journal}{\bibinfo{title}{Site-specific propagation
			prediction for wireless in-building personal communication system design}}.
	\newblock {\emph{\JournalTitle{IEEE transactions on Vehicular Technology}}}
	\textbf{\bibinfo{volume}{43}}, \bibinfo{pages}{879--891}
	(\bibinfo{year}{1994}).
	
	\bibitem{Gondarenko2008}
	\bibinfo{author}{Gondarenko, A.} \& \bibinfo{author}{Lipson, M.}
	\newblock \bibinfo{journal}{\bibinfo{title}{Low modal volume dipole-like
			dielectric slab resonator}}.
	\newblock {\emph{\JournalTitle{Optics express}}} \textbf{\bibinfo{volume}{16}},
	\bibinfo{pages}{17689--17694} (\bibinfo{year}{2008}).
	
	\bibitem{Hakansson2005}
	\bibinfo{author}{H{\aa}kansson, A.} \& \bibinfo{author}{S{\'a}nchez-Dehesa, J.}
	\newblock \bibinfo{journal}{\bibinfo{title}{Inverse designed photonic crystal
			de-multiplex waveguide coupler}}.
	\newblock {\emph{\JournalTitle{Optics Express}}} \textbf{\bibinfo{volume}{13}},
	\bibinfo{pages}{5440--5449} (\bibinfo{year}{2005}).
	
	\bibitem{ong2017freestanding}
	\bibinfo{author}{Ong, J.~R.}, \bibinfo{author}{Chu, H.~S.},
	\bibinfo{author}{Chen, V.~H.}, \bibinfo{author}{Zhu, A.~Y.} \&
	\bibinfo{author}{Genevet, P.}
	\newblock \bibinfo{journal}{\bibinfo{title}{Freestanding dielectric nanohole
			array metasurface for mid-infrared wavelength applications}}.
	\newblock {\emph{\JournalTitle{Optics letters}}} \textbf{\bibinfo{volume}{42}},
	\bibinfo{pages}{2639--2642} (\bibinfo{year}{2017}).
	
	\bibitem{Piggott2017}
	\bibinfo{author}{Piggott, A.~Y.}, \bibinfo{author}{Petykiewicz, J.},
	\bibinfo{author}{Su, L.} \& \bibinfo{author}{Vu{\v{c}}kovi{\'c}, J.}
	\newblock \bibinfo{journal}{\bibinfo{title}{Fabrication-constrained
			nanophotonic inverse design}}.
	\newblock {\emph{\JournalTitle{Scientific Reports}}}
	\textbf{\bibinfo{volume}{7}}, \bibinfo{pages}{1786} (\bibinfo{year}{2017}).
	
	\bibitem{Lu2013}
	\bibinfo{author}{Lu, J.} \& \bibinfo{author}{Vu{\v{c}}kovi{\'c}, J.}
	\newblock \bibinfo{journal}{\bibinfo{title}{Nanophotonic computational
			design}}.
	\newblock {\emph{\JournalTitle{Optics express}}} \textbf{\bibinfo{volume}{21}},
	\bibinfo{pages}{13351--13367} (\bibinfo{year}{2013}).
	
	\bibitem{Su2018}
	\bibinfo{author}{Su, L.}, \bibinfo{author}{Piggott, A.~Y.},
	\bibinfo{author}{Sapra, N.~V.}, \bibinfo{author}{Petykiewicz, J.} \&
	\bibinfo{author}{Vuckovic, J.}
	\newblock \bibinfo{journal}{\bibinfo{title}{Inverse design and demonstration of
			a compact on-chip narrowband three-channel wavelength demultiplexer}}.
	\newblock {\emph{\JournalTitle{ACS Photonics}}} \textbf{\bibinfo{volume}{5}},
	\bibinfo{pages}{301--305} (\bibinfo{year}{2017}).
	
	\bibitem{Frellsen2016}
	\bibinfo{author}{Frellsen, L.~F.}, \bibinfo{author}{Ding, Y.},
	\bibinfo{author}{Sigmund, O.} \& \bibinfo{author}{Frandsen, L.~H.}
	\newblock \bibinfo{journal}{\bibinfo{title}{Topology optimized mode
			multiplexing in silicon-on-insulator photonic wire waveguides}}.
	\newblock {\emph{\JournalTitle{Optics express}}} \textbf{\bibinfo{volume}{24}},
	\bibinfo{pages}{16866--16873} (\bibinfo{year}{2016}).
	
	\bibitem{Piggott2014}
	\bibinfo{author}{Piggott, A.~Y.} \emph{et~al.}
	\newblock \bibinfo{journal}{\bibinfo{title}{Inverse design and implementation
			of a wavelength demultiplexing grating coupler}}.
	\newblock {\emph{\JournalTitle{Scientific reports}}}
	\textbf{\bibinfo{volume}{4}}, \bibinfo{pages}{7210} (\bibinfo{year}{2014}).
	
	\bibitem{Englund2005}
	\bibinfo{author}{Englund, D.}, \bibinfo{author}{Fushman, I.} \&
	\bibinfo{author}{Vuckovic, J.}
	\newblock \bibinfo{journal}{\bibinfo{title}{General recipe for designing
			photonic crystal cavities}}.
	\newblock {\emph{\JournalTitle{Optics express}}} \textbf{\bibinfo{volume}{13}},
	\bibinfo{pages}{5961--5975} (\bibinfo{year}{2005}).
	
	\bibitem{molesky2018inverse}
	\bibinfo{author}{Molesky, S.} \emph{et~al.}
	\newblock \bibinfo{journal}{\bibinfo{title}{Inverse design in nanophotonics}}.
	\newblock {\emph{\JournalTitle{Nature Photonics}}}
	\textbf{\bibinfo{volume}{12}}, \bibinfo{pages}{659} (\bibinfo{year}{2018}).
	
	\bibitem{Mansouree:18}
	\bibinfo{author}{Mansouree, M.} \& \bibinfo{author}{Arbabi, A.}
	\newblock \bibinfo{title}{Large-scale metasurface design using the adjoint
		sensitivity technique}.
	\newblock In \emph{\bibinfo{booktitle}{Conference on Lasers and
			Electro-Optics}}, \bibinfo{pages}{FF1F.7},
	\doiprefix\url{10.1364/CLEO_QELS.2018.FF1F.7} (\bibinfo{publisher}{Optical
		Society of America}, \bibinfo{year}{2018}).
	
	\bibitem{melati2018mapping}
	\bibinfo{author}{Melati, D.} \emph{et~al.}
	\newblock \bibinfo{journal}{\bibinfo{title}{Mapping the global design space of
			integrated photonic components using machine learning pattern recognition}}.
	\newblock {\emph{\JournalTitle{arXiv preprint arXiv:1811.01048}}}
	(\bibinfo{year}{2018}).
	
	\bibitem{Ma2018}
	\bibinfo{author}{Ma, W.}, \bibinfo{author}{Cheng, F.} \& \bibinfo{author}{Liu,
		Y.}
	\newblock \bibinfo{journal}{\bibinfo{title}{Deep-learning enabled on-demand
			design of chiral metamaterials}}.
	\newblock {\emph{\JournalTitle{ACS nano}}}  (\bibinfo{year}{2018}).
	
	\bibitem{baxter2019plasmonic}
	\bibinfo{author}{Baxter, J.} \emph{et~al.}
	\newblock \bibinfo{journal}{\bibinfo{title}{Plasmonic colours predicted by deep
			learning}}.
	\newblock {\emph{\JournalTitle{arXiv preprint arXiv:1902.05898}}}
	(\bibinfo{year}{2019}).
	
	\bibitem{Liu2018a}
	\bibinfo{author}{Liu, D.}, \bibinfo{author}{Tan, Y.}, \bibinfo{author}{Khoram,
		E.} \& \bibinfo{author}{Yu, Z.}
	\newblock \bibinfo{journal}{\bibinfo{title}{Training deep neural networks for
			the inverse design of nanophotonic structures}}.
	\newblock {\emph{\JournalTitle{ACS Photonics}}} \textbf{\bibinfo{volume}{5}},
	\bibinfo{pages}{1365--1369} (\bibinfo{year}{2018}).
	
	\bibitem{Peurifoy2018}
	\bibinfo{author}{Peurifoy, J.} \emph{et~al.}
	\newblock \bibinfo{journal}{\bibinfo{title}{Nanophotonic particle simulation
			and inverse design using artificial neural networks}}.
	\newblock {\emph{\JournalTitle{Science advances}}}
	\textbf{\bibinfo{volume}{4}}, \bibinfo{pages}{eaar4206}
	(\bibinfo{year}{2018}).
	
	\bibitem{Liu2018}
	\bibinfo{author}{Liu, Z.}, \bibinfo{author}{Zhu, D.},
	\bibinfo{author}{Rodrigues, S.}, \bibinfo{author}{Lee, K.-T.} \&
	\bibinfo{author}{Cai, W.}
	\newblock \bibinfo{journal}{\bibinfo{title}{A generative model for the inverse
			design of metasurfaces}}.
	\newblock {\emph{\JournalTitle{Nano letters}}}  (\bibinfo{year}{2018}).
	
	\bibitem{Tahersima2018}
	\bibinfo{author}{Tahersima, M.~H.} \emph{et~al.}
	\newblock \bibinfo{journal}{\bibinfo{title}{Deep neural network inverse design
			of integrated nanophotonic devices}}.
	\newblock {\emph{\JournalTitle{arXiv preprint arXiv:1809.03555}}}
	(\bibinfo{year}{2018}).
	
	\bibitem{Zhang2018a}
	\bibinfo{author}{Zhang, T.} \emph{et~al.}
	\newblock \bibinfo{journal}{\bibinfo{title}{Spectrum prediction and inverse
			design for plasmonic waveguide system based on artificial neural networks}}.
	\newblock {\emph{\JournalTitle{arXiv preprint arXiv:1805.06410}}}
	(\bibinfo{year}{2018}).
	
	\bibitem{Qu2018}
	\bibinfo{author}{Qu, Y.}, \bibinfo{author}{Jing, L.}, \bibinfo{author}{Shen,
		Y.}, \bibinfo{author}{Qiu, M.} \& \bibinfo{author}{Soljacic, M.}
	\newblock \bibinfo{journal}{\bibinfo{title}{Migrating knowledge between
			physical scenarios based on artificial neural networks}}.
	\newblock {\emph{\JournalTitle{arXiv preprint arXiv:1809.00972}}}
	(\bibinfo{year}{2018}).
	
	\bibitem{Inampudi2018}
	\bibinfo{author}{Inampudi, S.} \& \bibinfo{author}{Mosallaei, H.}
	\newblock \bibinfo{journal}{\bibinfo{title}{Neural network based design of
			metagratings}}.
	\newblock {\emph{\JournalTitle{Applied Physics Letters}}}
	\textbf{\bibinfo{volume}{112}}, \bibinfo{pages}{241102}
	(\bibinfo{year}{2018}).
	
	\bibitem{yao2018intelligent}
	\bibinfo{author}{Yao, K.}, \bibinfo{author}{Unni, R.} \&
	\bibinfo{author}{Zheng, Y.}
	\newblock \bibinfo{journal}{\bibinfo{title}{Intelligent nanophotonics: merging
			photonics and artificial intelligence at the nanoscale}}.
	\newblock {\emph{\JournalTitle{arXiv preprint arXiv:1810.11709}}}
	(\bibinfo{year}{2018}).
	
	\bibitem{so2019simultaneous}
	\bibinfo{author}{So, S.}, \bibinfo{author}{Mun, J.} \& \bibinfo{author}{Rho,
		J.}
	\newblock \bibinfo{journal}{\bibinfo{title}{Simultaneous inverse design of
			materials and parameters of core-shell nanoparticle via deep-learning:
			Demonstration of dipole resonance engineering}}.
	\newblock {\emph{\JournalTitle{arXiv preprint arXiv:1904.02848}}}
	(\bibinfo{year}{2019}).
	
	\bibitem{asano2018optimization}
	\bibinfo{author}{Asano, T.} \& \bibinfo{author}{Noda, S.}
	\newblock \bibinfo{journal}{\bibinfo{title}{Optimization of photonic crystal
			nanocavities based on deep learning}}.
	\newblock {\emph{\JournalTitle{Optics express}}} \textbf{\bibinfo{volume}{26}},
	\bibinfo{pages}{32704--32717} (\bibinfo{year}{2018}).
	
	\bibitem{peurifoy2018nanophotonic}
	\bibinfo{author}{Peurifoy, J.} \emph{et~al.}
	\newblock \bibinfo{journal}{\bibinfo{title}{Nanophotonic particle simulation
			and inverse design using artificial neural networks}}.
	\newblock {\emph{\JournalTitle{Science advances}}}
	\textbf{\bibinfo{volume}{4}}, \bibinfo{pages}{eaar4206}
	(\bibinfo{year}{2018}).
	
	\bibitem{liu2018generative}
	\bibinfo{author}{Liu, Z.}, \bibinfo{author}{Zhu, D.},
	\bibinfo{author}{Rodrigues, S.~P.}, \bibinfo{author}{Lee, K.-T.} \&
	\bibinfo{author}{Cai, W.}
	\newblock \bibinfo{journal}{\bibinfo{title}{Generative model for the inverse
			design of metasurfaces}}.
	\newblock {\emph{\JournalTitle{Nano letters}}} \textbf{\bibinfo{volume}{18}},
	\bibinfo{pages}{6570--6576} (\bibinfo{year}{2018}).
	
	\bibitem{qu2018migrating}
	\bibinfo{author}{Qu, Y.}, \bibinfo{author}{Jing, L.}, \bibinfo{author}{Shen,
		Y.}, \bibinfo{author}{Qiu, M.} \& \bibinfo{author}{Soljacic, M.}
	\newblock \bibinfo{journal}{\bibinfo{title}{Migrating knowledge between
			physical scenarios based on artificial neural networks}}.
	\newblock {\emph{\JournalTitle{arXiv preprint arXiv:1809.00972}}}
	(\bibinfo{year}{2018}).
	
	\bibitem{pearson1901liii}
	\bibinfo{author}{Pearson, K.}
	\newblock \bibinfo{journal}{\bibinfo{title}{Liii. on lines and planes of
			closest fit to systems of points in space}}.
	\newblock {\emph{\JournalTitle{The London, Edinburgh, and Dublin Philosophical
				Magazine and Journal of Science}}} \textbf{\bibinfo{volume}{2}},
	\bibinfo{pages}{559--572} (\bibinfo{year}{1901}).
	
	\bibitem{scholkopf1997kernel}
	\bibinfo{author}{Sch{\"o}lkopf, B.}, \bibinfo{author}{Smola, A.} \&
	\bibinfo{author}{M{\"u}ller, K.-R.}
	\newblock \bibinfo{title}{Kernel principal component analysis}.
	\newblock In \emph{\bibinfo{booktitle}{International conference on artificial
			neural networks}}, \bibinfo{pages}{583--588}
	(\bibinfo{organization}{Springer}, \bibinfo{year}{1997}).
	
	\bibitem{Hinton2006}
	\bibinfo{author}{Hinton, G.~E.} \& \bibinfo{author}{Salakhutdinov, R.~R.}
	\newblock \bibinfo{journal}{\bibinfo{title}{Reducing the dimensionality of data
			with neural networks}}.
	\newblock {\emph{\JournalTitle{science}}} \textbf{\bibinfo{volume}{313}},
	\bibinfo{pages}{504--507} (\bibinfo{year}{2006}).
	
	\bibitem{friedman2001elements}
	\bibinfo{author}{Friedman, J.}, \bibinfo{author}{Hastie, T.} \&
	\bibinfo{author}{Tibshirani, R.}
	\newblock \emph{\bibinfo{title}{The elements of statistical learning}},
	vol.~\bibinfo{volume}{1} (\bibinfo{publisher}{Springer series in statistics
		New York}, \bibinfo{year}{2001}).
	
	\bibitem{kiarashinejad2019deep}
	\bibinfo{author}{Kiarashinejad, Y.}, \bibinfo{author}{Abdollahramezani, S.} \&
	\bibinfo{author}{Adibi, A.}
	\newblock \bibinfo{journal}{\bibinfo{title}{Deep learning approach based on
			dimensionality reduction for designing electromagnetic nanostructures}}.
	\newblock {\emph{\JournalTitle{arXiv preprint arXiv:1902.03865}}}
	(\bibinfo{year}{2019}).
	
	\bibitem{shportko2008resonant}
	\bibinfo{author}{Shportko, K.} \emph{et~al.}
	\newblock \bibinfo{journal}{\bibinfo{title}{Resonant bonding in crystalline
			phase-change materials}}.
	\newblock {\emph{\JournalTitle{Nature materials}}}
	\textbf{\bibinfo{volume}{7}}, \bibinfo{pages}{653} (\bibinfo{year}{2008}).
	
	\bibitem{abdollahramezani2018reconfigurable}
	\bibinfo{author}{Abdollahramezani, S.} \emph{et~al.}
	\newblock \bibinfo{journal}{\bibinfo{title}{Reconfigurable multifunctional
			metasurfaces employing hybrid phase-change plasmonic architecture}}.
	\newblock {\emph{\JournalTitle{arXiv preprint arXiv:1809.08907}}}
	(\bibinfo{year}{2018}).
	
\end{thebibliography}

\clearpage
\newpage

\newcommand{\beginsupplement}{%
	\renewcommand{\thetable}{S\arabic{table}}%
	\renewcommand{\theequation}{S\arabic{equation}}%
	\renewcommand{\thefigure}{S\arabic{figure}}%
	\renewcommand{\thesection}{S\arabic{section}}%
	\renewcommand{\thesubsection}{S\arabic{section}.\arabic{subsection}}%
}

\setcounter{figure}{0}
\setcounter{equation}{0}
\setcounter{section}{0}
\setcounter{subsection}{0}
\setcounter{table}{0}


\beginsupplement

\section*{Supplementary Information}

\subsection*{S1 Principal Component Analysis (PCA)}

PCA is a linear dimensionality reduction method, which maps the data into a linear subspace so that the variance is maximized. In other words, PCA projects data points onto the eigenvectors of the covariance matrix of the data points with larger eigenvalues. PCA results in minimum MSE during reconstruction. Thus, it gives the best representation of the data in the lower-dimensional space in terms of MSE. Considering $X$ as the response space matrix whose rows are samples of the reflection amplitude for an specific design, the first step in this algorithm is to centralize the dataset (i.e., subtract mean from data points): 

\begin{equation}
    \hat{X}=X-X_\textrm{mean},
\end{equation}
where $X_\textrm{mean}$ is the mean matrix of the data points, and $\hat{X}$ is the centralized matrix. The eigenvectors of the covariance matrix of $\hat{X}$ represent the principal components. These vectors are found using the singular value decomposition as: 

\begin{equation}
    \hat{X}=U\Sigma V^T,
\end{equation}

Here, columns of matrix $U$ represent the basis vectors of the covariance matrix $\hat{X}^T\hat{X}$. In addition, $\Sigma$ is a diagonal matrix, and its diagonal elements ($\sigma$) are the singular values associated with the columns of $U$. Therefore, by keeping the first $k$ columns of $U$ with the largest singular values, the projected matrix is:

\begin{equation}
    P\hat{X}=U_d^T \hat{X},
\end{equation}
where $U_d$ contains the first $d$ columns of $U$ and $P\hat{X}$ is the projection of the data in the lower-dimensional space. To reconstruct the projected response, we first make the projection inverse and then add the mean to the matrix as: 

\begin{equation}
    RX=U_dP\hat{X}+X_{mean},
\end{equation}
Finally, the reconstruction error is
\begin{equation}
    \frac{1}{N} \sum_{i=1}^{N} ||X_i-RX_i||_2^2,
\end{equation}
where $N$ is the number of data points, and $X_i$ represents the $i^\textrm{th}$ row of the response-space matrix.

\subsection*{S2 Kernel PCA (KPCA)}

\begin{figure}
    \centering
    \includegraphics[width=0.8\textwidth,trim={0cm 5cm 2cm 0cm},clip]{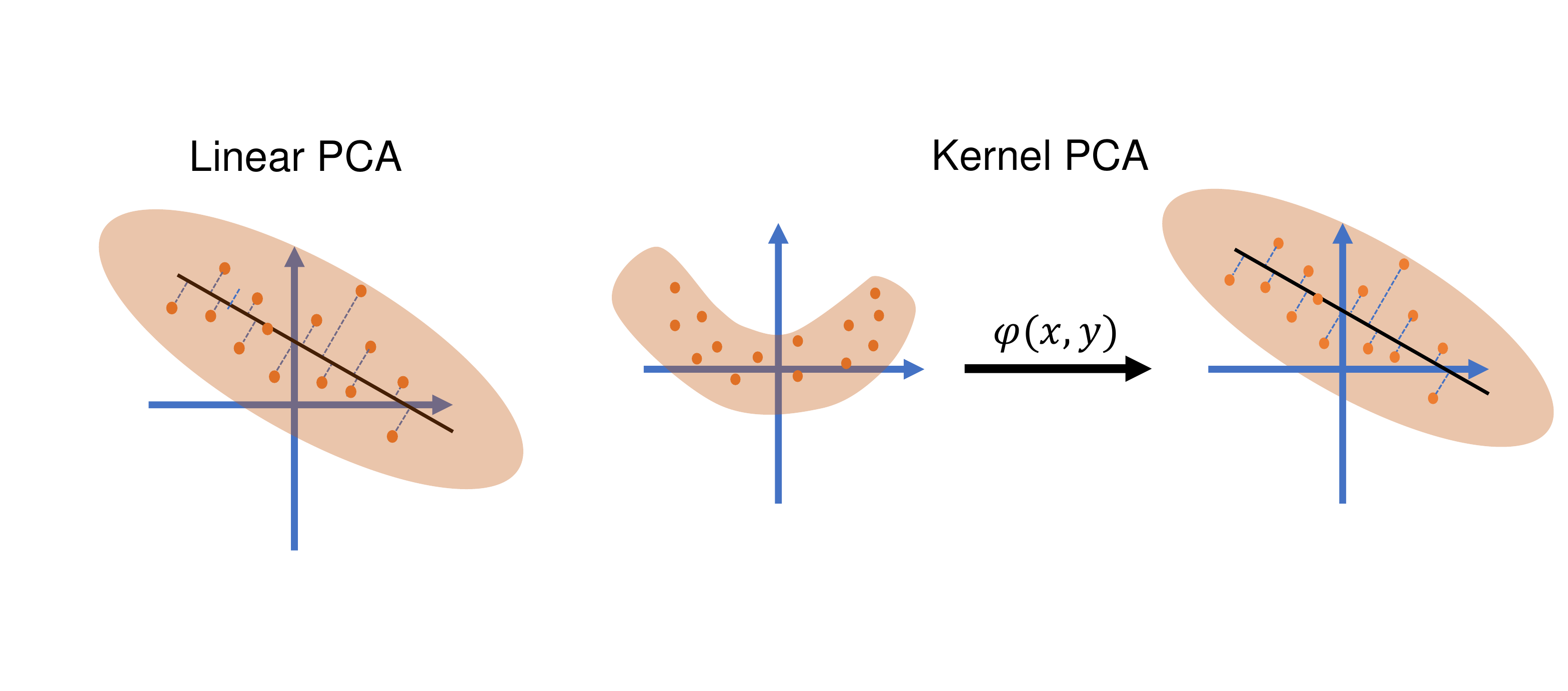}
    \caption{Linear PCA projects the data points onto the direction of largest principal component. Kernel PCA, however, maps the data points into another space using a nonlinear kernel function and then projects them on principal components of the new space.}
    \label{fig:fig13_PCA}
\end{figure}

The nonlinear version of PCA is known as kernel PCA (KPCA). Figure \ref{fig:fig13_PCA} shows how PCA and KPCA project data points into the lower-dimensional space. As we discussed before, PCA projects the data points on a linear subspace. However, if there are nonlinear properties in the dataset, PCA might result in a poor performance. KPCA transforms the original data into a higher-dimensional space using a nonlinear mapping $\phi(x_i)$ for all data points and then projects the transformed data into the lower-dimensional space. KPCA provides better results when we are interested in nonlinear relation in the dataset. 

The kernel function is defined as $k(x_i,x_j)=\phi(x_i)\phi(x_j )^T$. Two well-known kernels are polynomial kernel $k(x_i,k_j)=(x_i x_j^T+c)^m$ and Gaussian Kernel $k(x_i,k_j)=e^{-||x_i- x_j||_2^2/2\gamma}$, where $m$ and $\gamma$ are the free parameters. The best parameters then could be found using the cross-validation technique. In this work, we used the polynomial kernel to reduce the dimension of the data and compared the results with the other methods. To implement the kernel PCA, we should do the following:

\begin{itemize}
    \item Compute the Gram Matrix K where $K_{ij}=k(x_i,x_j)$. Note that the dataset is centralized and has zero mean. The Gram Matrix is as below:
    \begin{equation}
        \hat{K}=K-\textbf{1}_NK-K\textbf{1}_N+\textbf{1}_NK\textbf{1}_N
    \end{equation}
    where $\textbf{1}_N$ is a $N \times N$ matrix with all elements equal to $1/N$.  
    \item Find the basis vectors of the transformed space by using eigen-decomposition of the Gram Matrix. 
    \item Project the data points on the first $d$ eigenvectors with higher eigenvalues.
\end{itemize}

\subsection*{S3 Autoencoder}
Autoencoder is a neural-based dimensionality reduction network. Figure \ref{fig:fig2_AE} shows a schematic of an autoencoder with a multilayered NN to map the high-dimensional input on the leftmost layer to the low-dimensional data in the middle layer. The same NN can be used to decode and recover the data back to the original space with a specific error. Actually, the data from the input layer are first compressed and subsequently are uncompressed into those closely matches the original data. For the simplest case(i.e. mono-layer encoder and mono-layer decoder) high-dimensional and low-dimensional data relation are:
\begin{equation}
    z=\sigma(Wx+b),
\end{equation}
\begin{equation}
    X'=\sigma'(W'x+b'),
\end{equation}
Which $x$, $z$, $W$, $W'$, $b$, $b'$, $\sigma$, and $\sigma'$ represent high-dimensional data, low-dimensional data, encoder weight matrix, decoder weight matrix, encoder bias, decoder bias, encoder activation function, decoder activation function respectively .To find the optimum weights for autoencoder
The MSE loss function should be minimized:\\
\begin{equation}
    L=||X-X'||^2=||X-\sigma'(W'(\sigma(Wx+b))+b')||^2,
\end{equation}
\\
\subsection*{S4 Comparison of PCA, KPCA, and Autoencoder}
\begin{figure}[b]
\centering
\begin{subfigure}{}
    \includegraphics[page=1,width=0.4\textwidth, trim={3cm 11cm 0cm 3cm},clip]{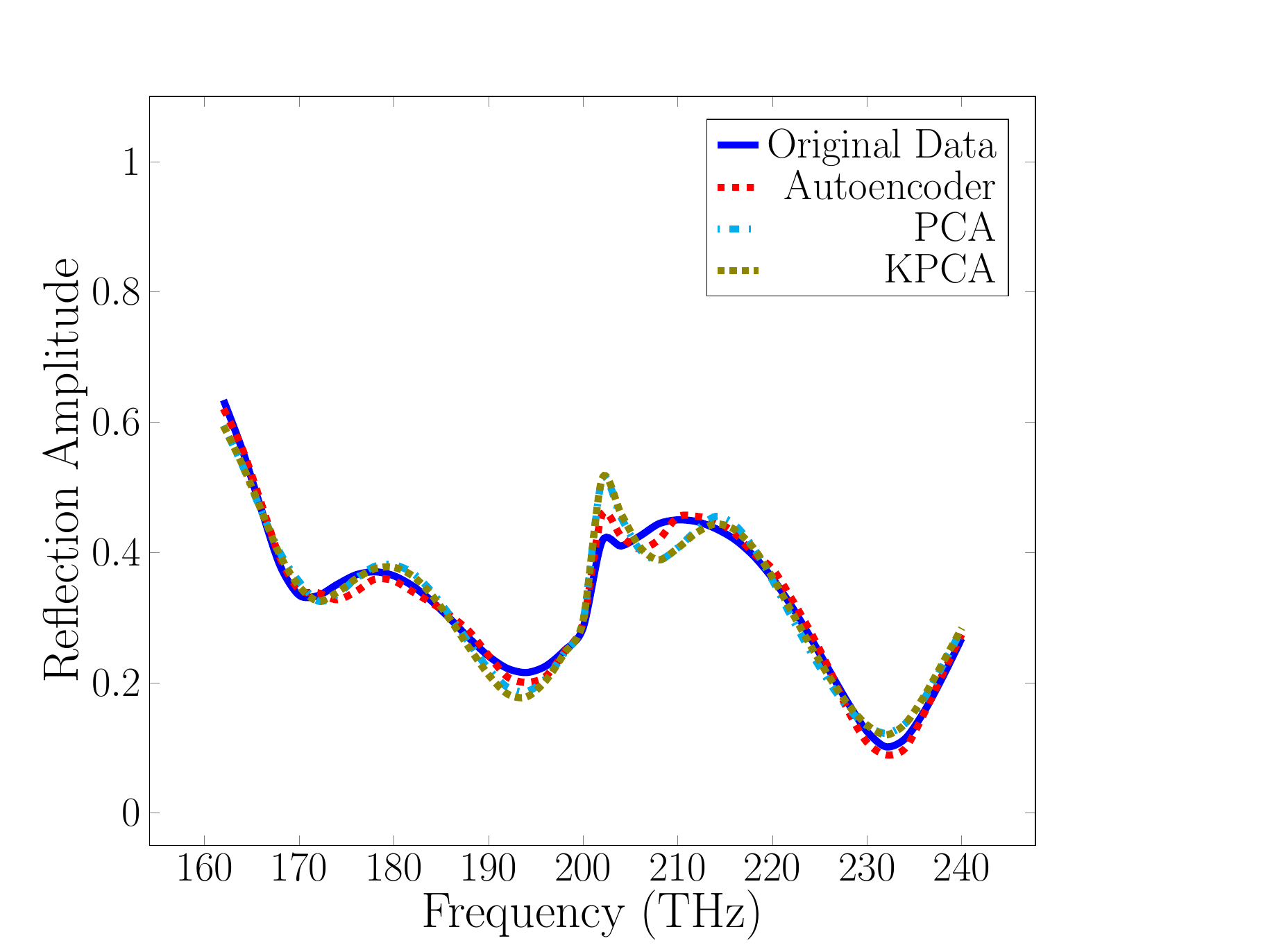}
    \includegraphics[page=2,width=0.4\textwidth, trim={3cm 11cm 0cm 3cm},clip]{Fig_Rec_Final.pdf}
\end{subfigure}%

\begin{subfigure}{}
    \includegraphics[page=3,width=0.4\textwidth, trim={3cm 11cm 0cm 3cm},clip]{Fig_Rec_Final.pdf}
    \includegraphics[page=4,width=0.4\textwidth, trim={3cm 11cm 0cm 3cm},clip]{Fig_Rec_Final.pdf}
\end{subfigure}

\begin{subfigure}{}
    \includegraphics[page=5,width=0.4\textwidth, trim={3cm 11cm 0cm 3cm},clip]{Fig_Rec_Final.pdf}
    \includegraphics[page=6,width=0.4\textwidth, trim={3cm 11cm 0cm 3cm},clip]{Fig_Rec_Final.pdf}
\end{subfigure}

\begin{subfigure}{}
    \includegraphics[page=7,width=0.4\textwidth, trim={3cm 11cm 0cm 3cm},clip]{Fig_Rec_Final.pdf}
    \includegraphics[page=8,width=0.4\textwidth, trim={3cm 11cm 0cm 3cm},clip]{Fig_Rec_Final.pdf}
   
\end{subfigure}

\caption{Comparison of the reflection spectra of the original and reconstructed data after reducing the dimensionality of the response space employing different methods (PCA, KPCA, and autoencoder) with $d$ = 7.}
    \label{fig:fig15_RecAll}
\end{figure}

\begin{figure}[b]
\centering
\begin{subfigure}{}
    \includegraphics[page=9,width=0.4\textwidth, trim={3cm 11cm 0cm 3cm},clip]{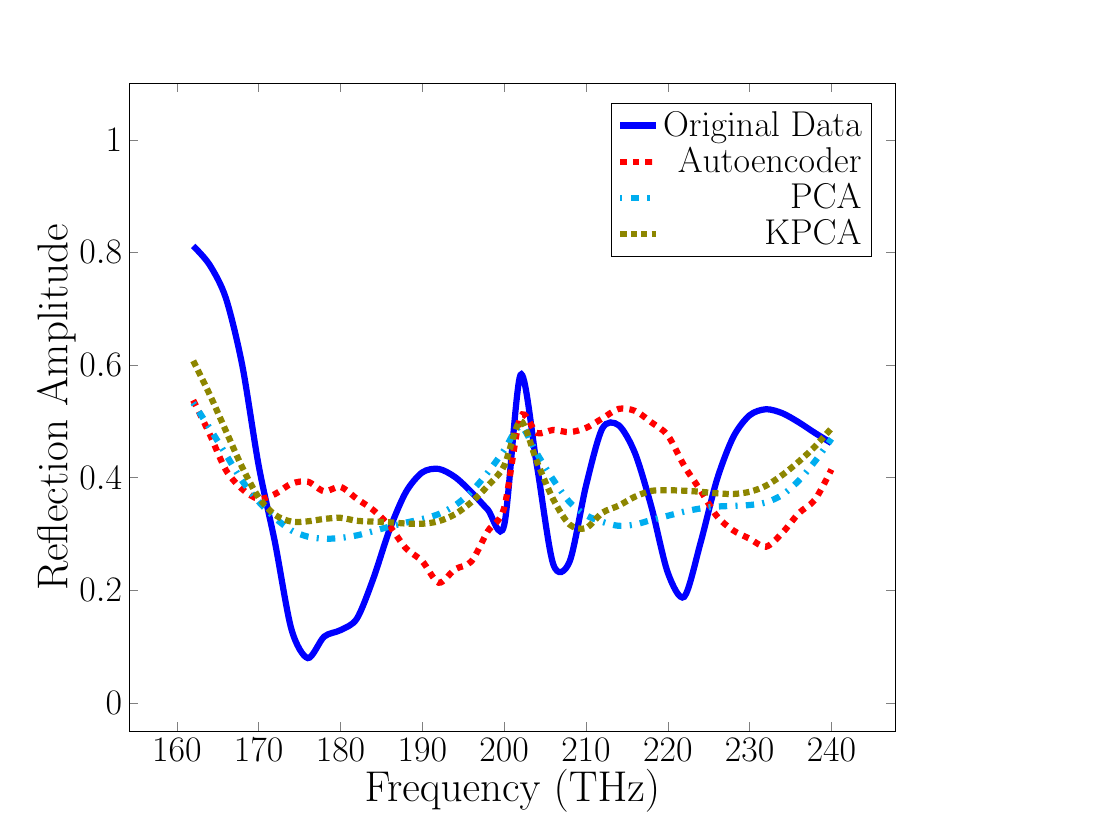}
    \includegraphics[page=10,width=0.4\textwidth, trim={3cm 11cm 0cm 3cm},clip]{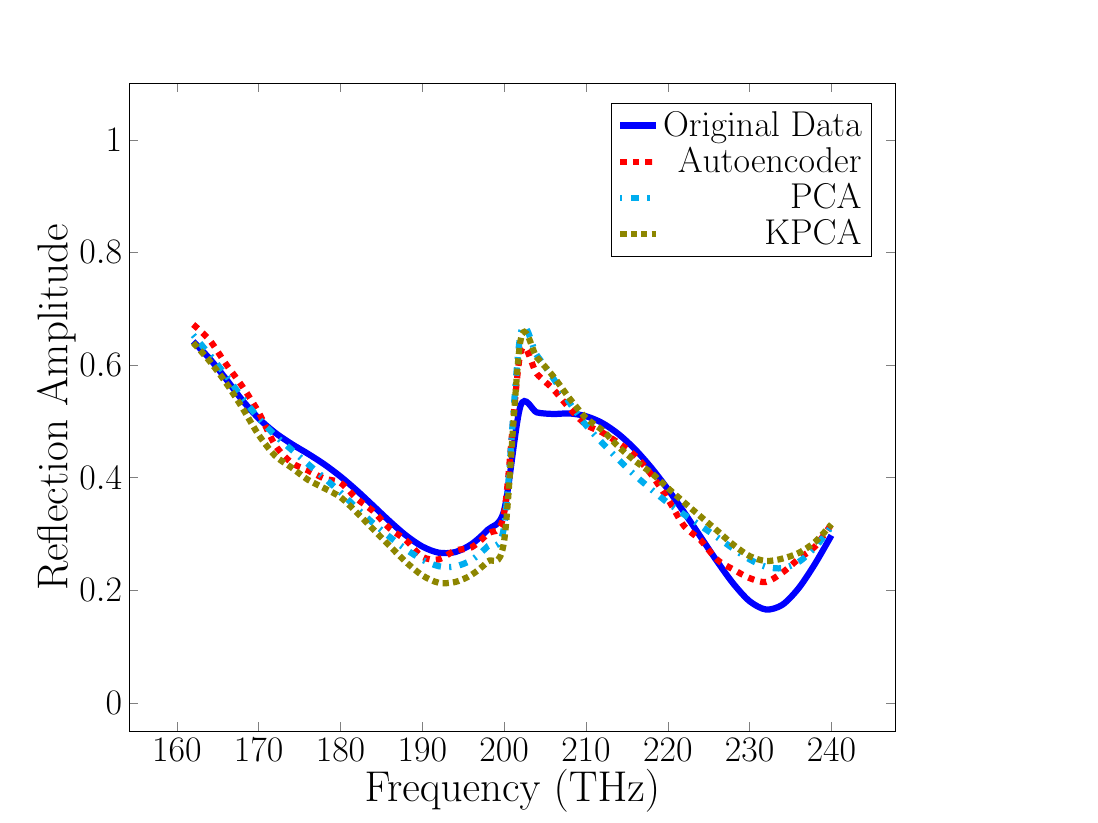}
\end{subfigure}%

\begin{subfigure}{}
    \includegraphics[page=11,width=0.4\textwidth, trim={3cm 11cm 0cm 3cm},clip]{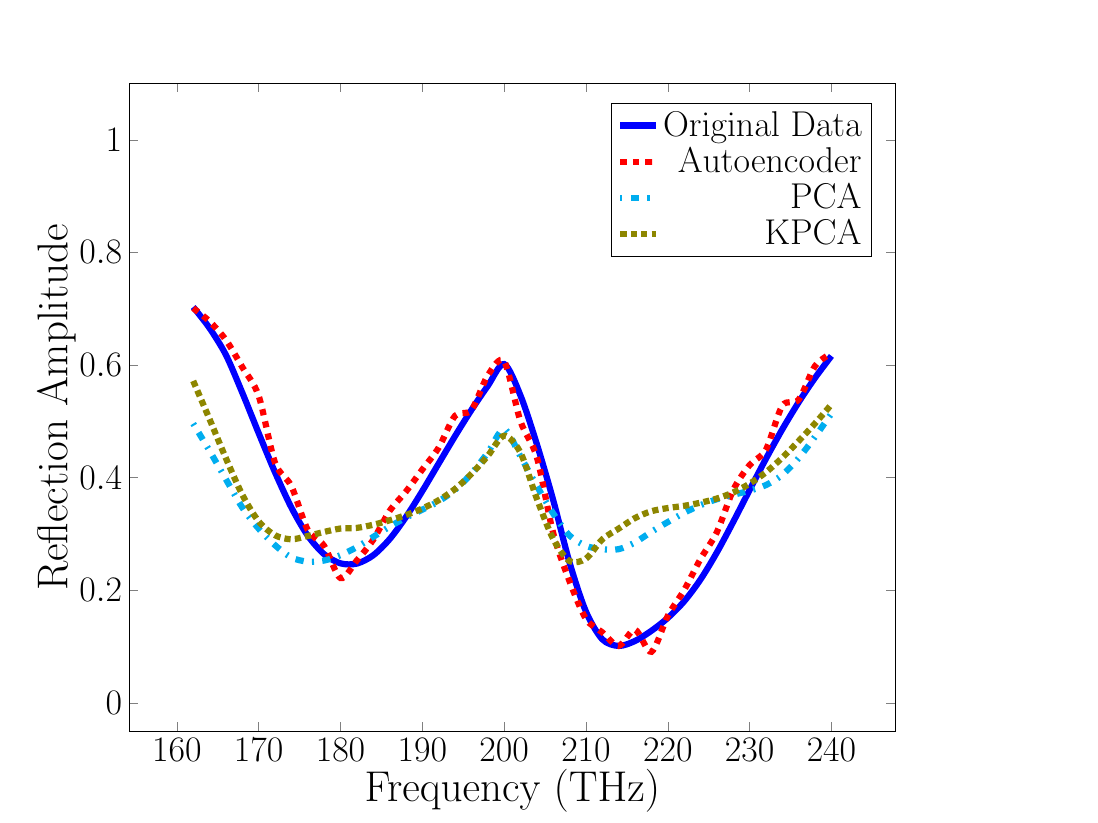}
    \includegraphics[page=12,width=0.4\textwidth, trim={3cm 11cm 0cm 3cm},clip]{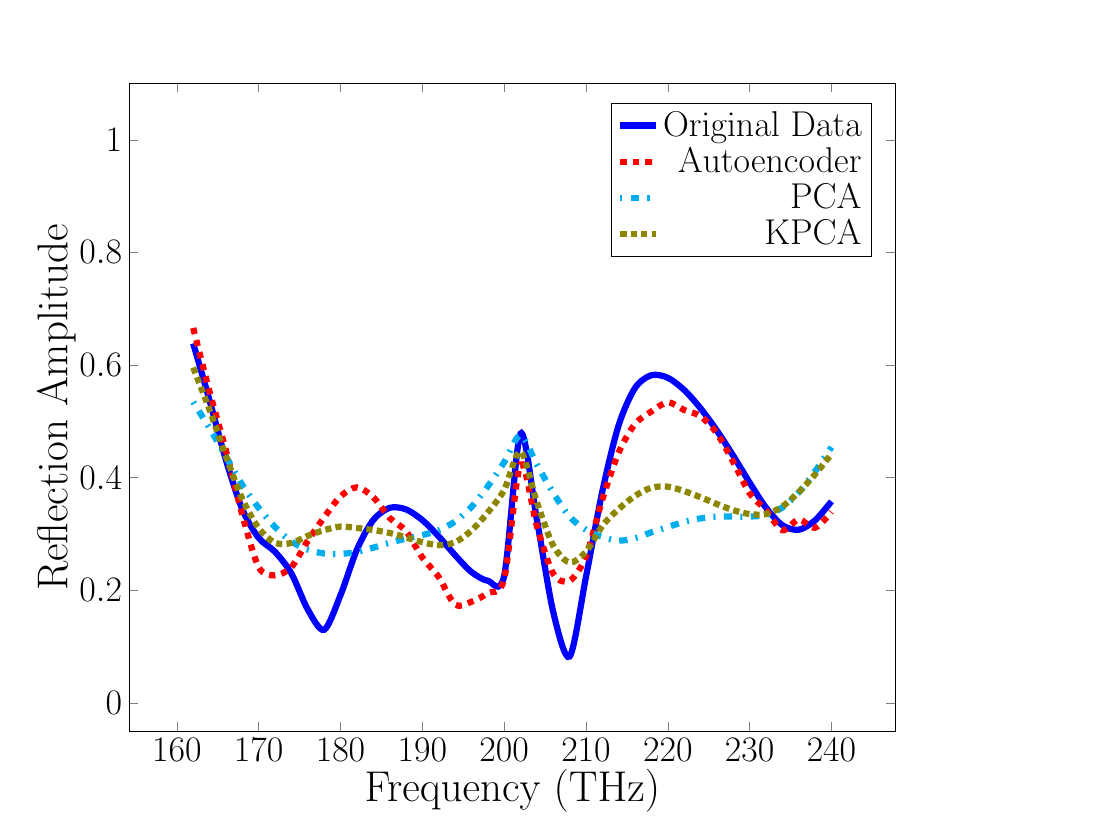}
\end{subfigure}

\begin{subfigure}{}
    \includegraphics[page=13,width=0.4\textwidth, trim={3cm 11cm 0cm 3cm},clip]{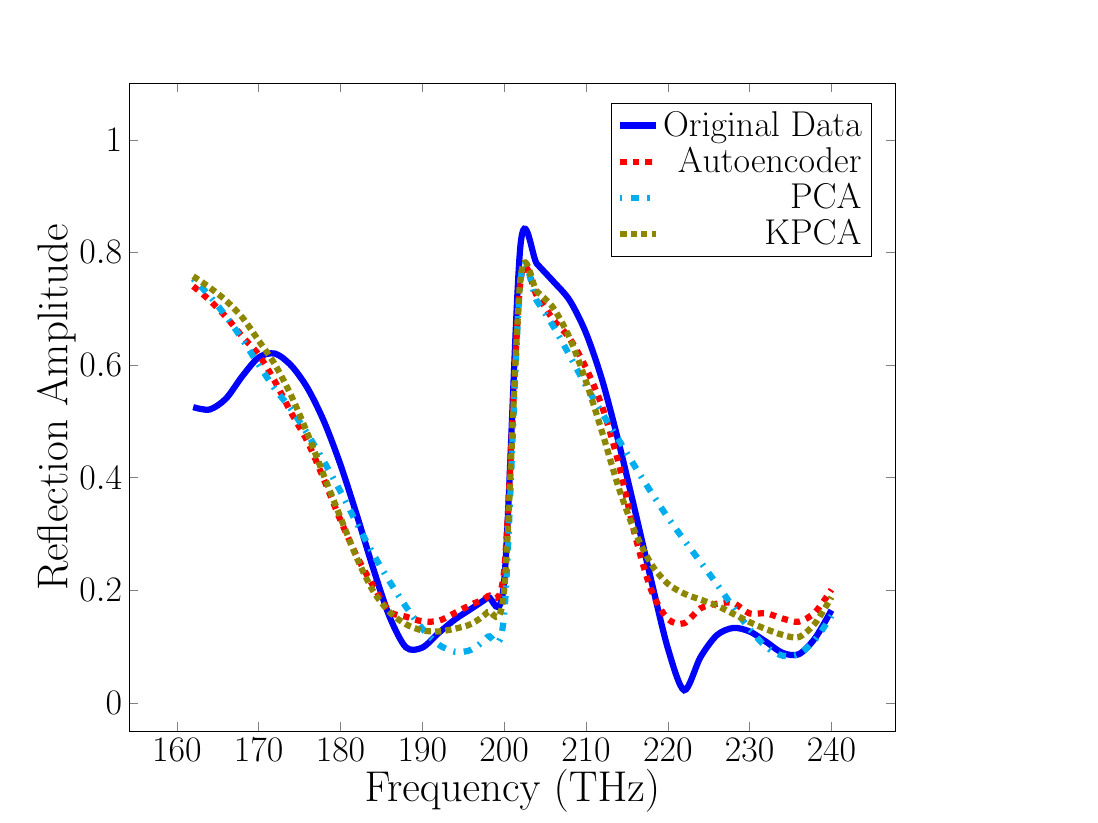}
    \includegraphics[page=14,width=0.4\textwidth, trim={3cm 11cm 0cm 3cm},clip]{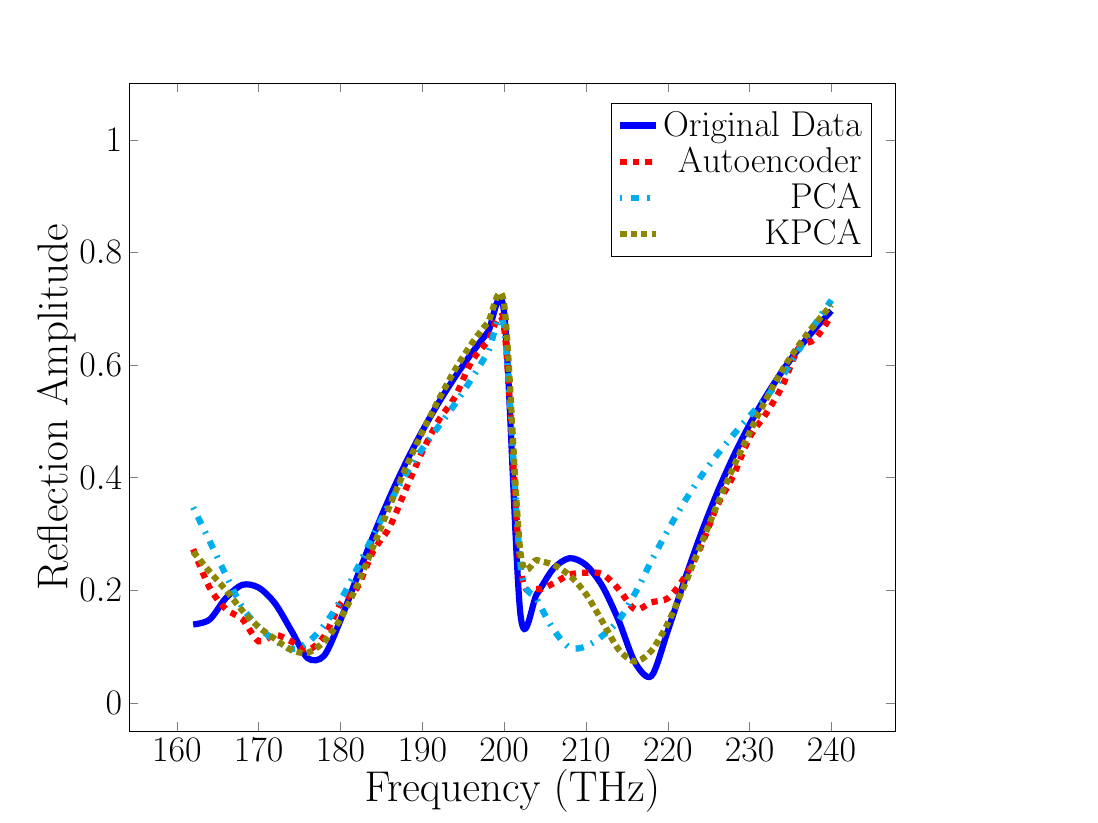}
\end{subfigure}

\begin{subfigure}{}
    \includegraphics[page=15,width=0.4\textwidth, trim={3cm 11cm 0cm 3cm},clip]{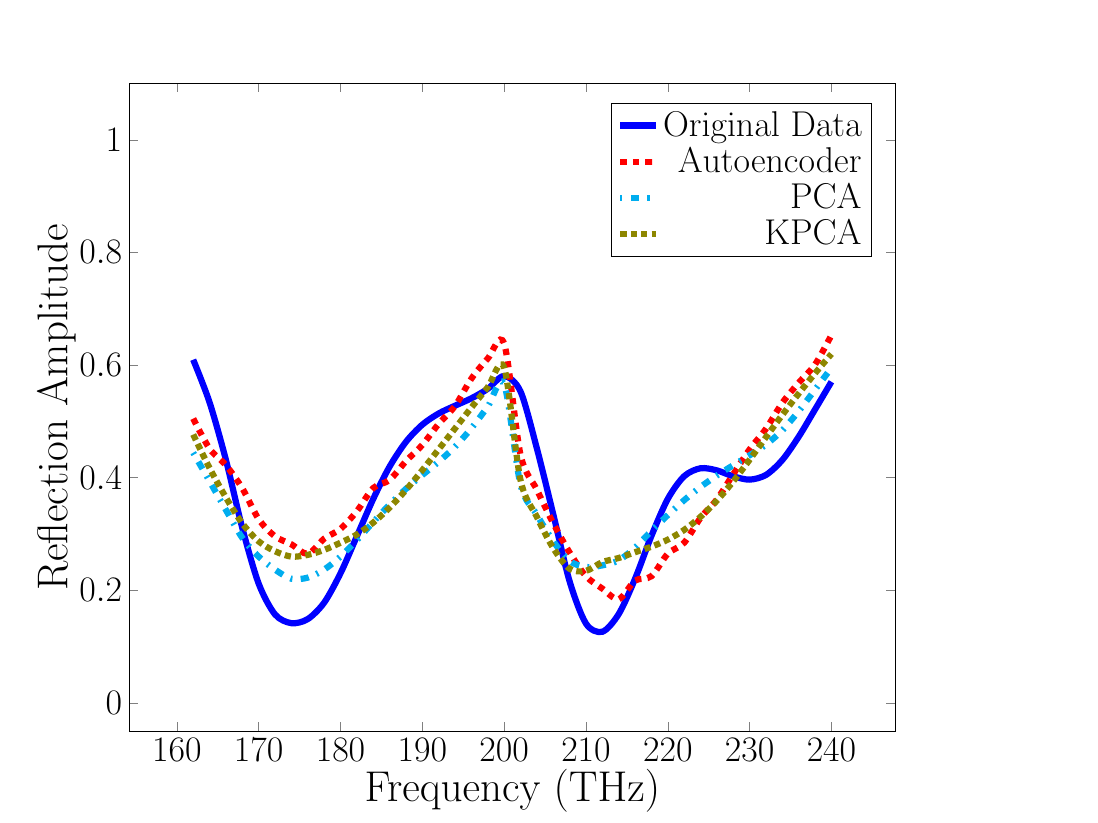}
    \includegraphics[page=16,width=0.4\textwidth, trim={3cm 11cm 0cm 3cm},clip]{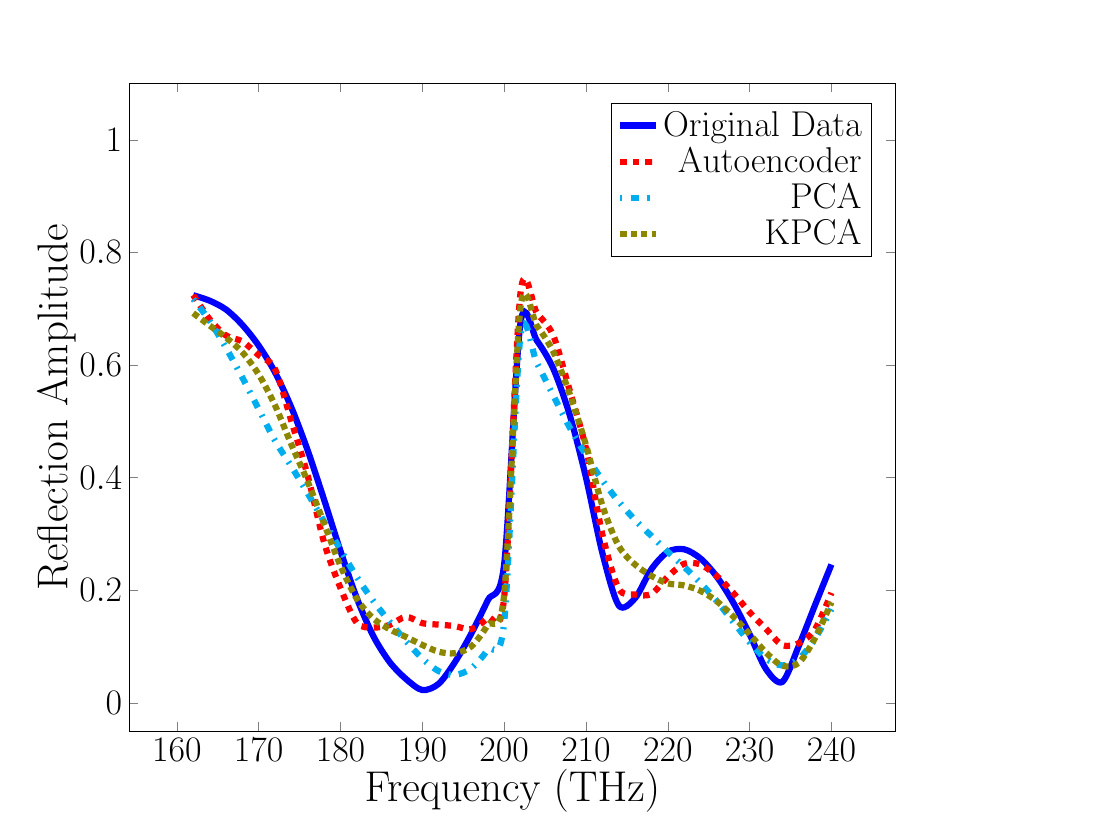}
   
\end{subfigure}

\caption{Comparison of the reflection spectra of the original and reconstructed data after reducing the dimensionality of the response space employing different methods (PCA, KPCA, and autoencoder) with $d$ = 2.}
    \label{fig:fig16_RecAll}
\end{figure}

Figure \ref{fig:fig15_RecAll} and \ref{fig:fig16_RecAll} demonstrate the performance of PCA, KPCA, and autoencoder for reconstructing the reflection spectra from the reduced space. The results represent the effectiveness of the dimensionality reduction in recovering reflection spectra after finding the optimum low-dimensional space.

\end{document}